\documentclass[aps,preprint,prd,superscriptaddress,floatfix, nofootinbib]{revtex4}

\usepackage{amsmath,amssymb}
\usepackage{cancel}
\newcommand{\ua}{$U(1)_A$ } 
\newcommand{\su}{$SU(2)_L\times SU(2)_R$ }
\usepackage{xspace}
\newcommand{\mobius}{M\"obius\xspace}
\newcommand{\gw}{Ginsparg-Wilson relation\xspace}
\newcommand{\av}[1]{\langle #1 \rangle} 
\newcommand{\tr}{\mathrm{tr}\,}  

\usepackage{graphicx}
\usepackage{color}

\usepackage{ifpdf}
\ifpdf
  \usepackage{epstopdf}
\fi

\newcommand{\dsection}[1]{}
\renewcommand{\include}[1]{}

\begin{document}

\preprint{OU-HET-905, KEK-CP-349, YITP-16-108}

\title{Evidence of effective axial $U(1)$ symmetry restoration at high temperature QCD}


\author{A.~Tomiya}
\affiliation{
  Key Laboratory of Quark \& Lepton Physics (MOE) and Institute of
  Particle Physics, 
  Central China Normal University, Wuhan 430079, China} 
\author{G.~Cossu}
\affiliation{
School of Physics and Astronomy, The University of Edinburgh, Edinburgh EH9 3JZ, United Kingdom 
}
\author{S.~Aoki}
\affiliation{
  Center for Gravitational Physics, Yukawa Institute for Theoretical Physics, 
  Kyoto University, 
  Kyoto 606-8502, Japan}
\author{H.~Fukaya}
\affiliation{
  Department of Physics, Graduate School of Science, 
  Osaka University,
  Toyonaka 560-0043, Japan} 
\author{S.~Hashimoto}
\affiliation{
  High Energy Accelerator Research Organization (KEK), 
  Tsukuba 305-0801, Japan}
\affiliation{
  School of High Energy Accelerator Science, 
  The Graduate University for Advanced Studies (Sokendai), 
  Tsukuba 305-0801, Japan}
\author{T.~Kaneko}
\affiliation{
  High Energy Accelerator Research Organization (KEK), 
  Tsukuba 305-0801, Japan}
\affiliation{
  School of High Energy Accelerator Science, 
  The Graduate University for Advanced Studies (Sokendai), 
  Tsukuba 305-0801, Japan}
\author{J.~Noaki}
\affiliation{
  High Energy Accelerator Research Organization (KEK), 
  Tsukuba 305-0801, Japan}

\collaboration{JLQCD collaboration}
\noaffiliation

\begin{abstract}
We study the axial $U(1)$ symmetry at finite temperature in two-flavor lattice QCD.
Employing the \mobius domain-wall fermions, we generate gauge configurations slightly above
the critical temperature $T_c$ with different lattice sizes $L=$ 2--4 fm.
Our action allows frequent topology tunneling while keeping good chiral symmetry close enough
to that of overlap fermions.
This allows us to recover full chiral symmetry by an overlap/domain-wall reweighting.
Above the phase transition, a strong suppression of the low-lying modes is observed
in both of overlap and domain-wall Dirac spectra.
We, however, find a sizable violation of the Ginsparg-Wilson relation
in the \mobius domain-wall Dirac eigenmodes, which 
dominates the signals of the axial $U(1)$ symmetry breaking near the chiral limit.
We also find that the use of overlap fermion only in the valence sector is dangerous
since it suffers from the  artifacts due to partial quenching.
Reweighting the \mobius domain-wall fermion determinant to that of the overlap fermion,
we observe the axial U(1) breaking  to vanish in the chiral limit,
which is stable against the changes of the lattice volume and lattice spacing.
\end{abstract}


\maketitle


\section{Introduction}
The action of Quantum Chromodynamics (QCD) with two massless quark
flavors has a global
$SU(2)_L\times SU (2)_R\times U(1)_V\times U(1)_A$
symmetry.
The flavor (or isospin) non-singlet part $SU(2)_L\times SU(2)_R$ is
spontaneously broken to the vector-like sub-group $SU(2)_V$ 
below the critical temperature $T_c$ by the presence of 
the chiral condensate $\av{\bar\psi\psi} \neq 0$. 
The axial \ua symmetry is, on the other hand, violated by anomaly.
Namely, the flavor-singlet axial current is not conserved due to the 
topological charge density operator appearing in
the axial Ward-Takahashi identity.
Since this anomalous Ward-Takahashi identity is valid in any
environment, the \ua symmetry is supposed to be violated at any
temperature. 
Taking account of the gluonic dynamics, on the other hand, 
how much the topological charge density contributes to the low-energy physics 
may depend on the amount of topological activity in the
background gauge field.
In fact, at high temperature $T\gg T_c$ \cite{Gross:1980br}, 
the instanton density is exponentially suppressed and the \ua
symmetry, as probed by physical observables, would be restored.

Just above the transition temperature $T_c$, 
topological fluctuations are not well understood theoretically,
due to non-perturbative nature of QCD dynamics, and the question remains open
about whether the \ua symmetry is effectively restored or not.
It is related to the important question on the order and the critical
exponents of two-flavor QCD chiral phase transition, 
since the symmetry determines the properties of the transition as discussed in
\cite{Pisarski:1983ms}\footnote{
The original argument of \cite{Pisarski:1983ms} was based on the one-loop computation
of the effective meson theory. The order of the QCD chiral phase transition is a subject of active studies.
The recent developments are found in 
Refs.~\cite{Cossu:2007mn,Pelissetto:2013hqa,Ishikawa:2013tua,Nakayama:2014sba,Fejos:2014qga,Grahl:2014fna,Kanazawa:2014cua,Sato:2014axa,Meggiolaro:2014eua, Grahl:2014fna,Kanazawa:2015xna,Eser:2015pka,Ejiri:2015vip,Nakayama:2016jhq,Bonati:2014kpa}.
}.
The fate of the \ua symmetry is also of phenomenological interest,
since the topological susceptibility in the hot early universe
gives a strong constraint on the axion dark matter scenario \cite{Berkowitz:2015aua,Kitano:2015fla,Petreczky:2016vrs,Borsanyi:2016ksw,Bonati:2015vqz}.

One of the possible observables for the $U(1)_A$ symmetry breaking is the 
difference of flavor non-singlet meson susceptibilities, 
\begin{equation}
  \label{eq:Delta_def}
  \Delta_{\pi-\delta} = 
\int d^4x
  \left[
    \langle\pi^a(x)\pi^a(0)\rangle -
    \langle\delta^a(x)\delta^a(0)\rangle
  \right],
\end{equation}
where $\pi^a = \bar{\psi}\tau^a\gamma_5\psi$ and $\delta^a = \bar{\psi}\tau^a\psi$ represent 
the isospin triplet pseudo-scalar and scalar operators, respectively.
Here $\tau^a$ denotes one of the $SU(2)$ generators. 
The measurement of (\ref{eq:Delta_def}) is relatively easy as it does not
involve disconnected diagrams.
Decomposing the quark propagator into the eigenmodes of
the Dirac operator, $\Delta_{\pi-\delta}$ may be written only in terms of its
eigenvalue spectrum $\rho(\lambda)$ (in the continuum limit):
\begin{equation}
  \label{eq:Delta_eig}
  \Delta_{\pi-\delta} = \int_0^\infty d\lambda\,\rho(\lambda)
  \frac{2m^2}{(\lambda^2+m^2)^2}.
\end{equation}
Here the Dirac operator eigenvalue density is defined by
$\rho(\lambda)=(1/V)\langle\sum_{\lambda'}\delta(\lambda-\lambda')\rangle$,
with the four-dimensional volume $V=L^3\times L_t$.
To be precise, this relation is valid only in the 
large volume limit $L^3 \to \infty$.
We need to take the spatial lattice size $L$ much larger than 
the correlation length of the system so that corrections of order $1/L$ are negligible.
The size $L_t$ in the temporal direction corresponds to the inverse temperature $1/T$.
Since the integrand of (\ref{eq:Delta_eig}) at finite eigenvalue 
$\lambda$ vanishes in the massless limit,
$\rho(0)$ (including its derivatives) controls whether $\Delta_{\pi-\delta}$ survives 
or not above the chiral phase transition temperature \cite{Cohen:1996ng, Cohen:1997hz, Birse:1996dx}.
In fact, under an assumption of analyticity (in $m^2$) it can be shown that 
$\rho(0)$ vanishes {\it before} taking the massless limit \cite{Aoki:2012yj},
and therefore the integral (\ref{eq:Delta_eig}) vanishes.
Our previous work using overlap fermions \cite{Cossu:2013uua}, as well as the work by the TWQCD
collaboration \cite{Chiu:2013wwa}, supports this argument.
We also note that a recent numerical simulation with Wilson fermions \cite{Brandt:2016daq}
and an analytic study \cite{Azcoiti:2016zbi} report that
the $U(1)_A$ anomaly effect is consistent with zero in the chiral limit.

A possible counter argument to \cite{Aoki:2012yj} is that the spectral
function could be non-analytic near $\lambda=0$ in the infinite volume limit.
Some of the recent lattice calculations suggest this possibility by
finding a peak of $\rho(\lambda)$ near $\lambda=0$
\cite{Bazavov:2012qja,Buchoff:2013nra,Dick:2015twa}.
There is however a delicate issue due to the violation of chiral
symmetry on the lattice, since the zero modes are necessarily chiral
(left- or right-handed) in the continuum QCD, and even tiny violation
of chiral symmetry in the lattice fermion formulation may induce
spurious zero modes or destroy the physical zero modes.
In fact, we demonstrated in \cite{Cossu:2015kfa} that the near-zero modes are largely affected
by the violation of the Ginsparg-Wilson relation \cite{Ginsparg:1981bj}\footnote{
We do not observe such large violation of chirality at low temperature. 
It seems that this problem arises only at $T>T_c$.
One can intuitively understand the qualitative difference between
$T>T_c$ and $T<T_c$ by the amount of the physical
near-zero modes, which survive in the continuum limit.
At $T<T_c$, there are dense physical modes mixing with lattice
artifact and make the effect of the violation of chiral symmetry relatively small,
while at $T>T_c$, the unphysical near-zero  may give a dominant effect on the low-energy observables.
} 
even if one uses the
M\"obius domain-wall fermion, which suppresses the violation of chiral
symmetry at the level of the residual mass being $O(1\mbox{~MeV})$.

In this work we study the spectral function of the Dirac operator at
finite temperature in order to elucidate the possible effect of the
residual chiral symmetry violation in the M\"obius domain-wall fermion formulation.
We introduce the reweighting technique to realize the formulation that
exactly satisfies the Ginsparg-Wilson relation based on the ensembles
generated with the M\"obius domain-wall fermions \cite{Brower:2005qw,Brower:2012vk}.
Any difference between the spectral functions before and after the
reweighting may indicate contamination due to imprecise chiral symmetry.
In our previous work \cite{Cossu:2013uua} we employed the overlap
fermion formulation \cite{Neuberger:1997fp, Neuberger:1998my}  that has exact chiral symmetry \cite{Luscher:1998pqa}, 
but the simulation was restricted to a fixed
topological sector \cite{Fukaya:2006vs, Aoki:2008tq}.
We avoided this problem in the present work by generating ensembles with
the M\"obius domain-wall fermion, that allows frequent
topology changes during simulations.

We generate ensembles of two-flavor QCD configurations at temperatures in the range
between 170 and 220~MeV, that cover the temperature region slightly
above the chiral phase transition.
We employ the M\"obius domain-wall
fermion for the sea quark formulation to achieve good chiral symmetry,
while allowing the topology tunnelings.
Spatial volume sizes are in the range $L$ = 2--4~fm.
Degenerate bare quark masses are taken in the range 2--25~MeV.

As we will see below, we find a strong suppression of 
low-lying Dirac eigenvalues and the results for
the \mobius domain-wall and overlap fermions agree with each other.
The \ua susceptibility $\Delta_{\pi-\delta}$ is more subtle.
It turns out that the violation of the Ginsparg-Wilson relation
for the lowest modes of \mobius domain-wall fermion
is much larger than what we expect from the residual mass $\sim O(1)$ MeV, 
and the violation dominates the signals of $\Delta_{\pi-\delta}$ \cite{Cossu:2015kfa}.
We also find that the use of overlap fermion only in the valence sector 
(as proposed in \cite{Dick:2015twa, Sharma:2016cmz})
suffers from partially quenching artifacts.
Replacing  the \mobius domain-wall Dirac operator 
by the overlap Dirac operator both in valence and sea quarks, 
we find 
a strong suppression of $\Delta_{\pi-\delta}$ towards the chiral limit.
The chiral extrapolation of $\Delta_{\pi-\delta}$ is consistent with
zero, which is insensitive to the change of lattice volume $V$ 
and lattice spacing $a$.

The rest of this paper is organized as follows. 
In Sec.~\ref{sec:setup}, we explain the technical details of our numerical setup.
The result for the Dirac eigenvalue spectrum is presented in Sec.~\ref{sec:Dirac}, 
and that for the \ua susceptibility is given in Sec.~\ref{sec:sus}.
Our conclusions are given in Sec.~\ref{sec:conclusion}.
Preliminary reports of this work can be  found in \cite{Cossu:2014aua, Tomiya:2014mma, Cossu:2015lnb, Aok:2016gxw}.

\section{Lattice setup}
\label{sec:setup}

\subsection{M\"obius domain-wall and overlap fermions}
In this work we employ the M\"obius domain-wall fermion
\cite{Brower:2005qw,Brower:2012vk}, which is numerically less expensive than
the overlap fermion and allows topology tunneling at the cost of
violating the Ginsparg-Wilson relation \cite{Ginsparg:1981bj} at some small amount, 
which is controllable by the finite size $L_s$ of the fifth direction.
This formulation is one of many possible implementations of lattice
fermions satisfying the Ginsparg-Wilson relation. 
They are classified by the kernel operator and the approximation of
the sign function.
As described below, the M\"obius domain-wall fermion has the same
Shamir kernel $D_W/(2+D_W)$ \cite{Shamir:1993zy,Furman:1994ky}, 
with $D_W$ the Wilson-Dirac operator, as that of the conventional domain-wall
fermion \cite{Kaplan:1992bt, Shamir:1993zy,Furman:1994ky}, 
while improving the approximation of the sign function by introducing a scale
parameter to the kernel.
In this sense, the M\"obius domain-wall fermion is a {\it better}
domain-wall fermion.
The overlap fermion formulation of Neuberger \cite{Neuberger:1997fp, Neuberger:1998my} has a different kernel,
$D_W$, and the rational approximation of the sign function is typically
adopted. 
As far as the chiral symmetry of the resulting fermion is
concerned, the difference of the kernels and the details of the
sign function approximation are not important.

In the following, we take the lattice spacing  $a=1$ unless otherwise stated.
It is shown that the fermion determinant generated with the
domain-wall fermion together with the associated Pauli-Villars field is
equivalent to a determinant of the four-dimensional (4D) effective
operator \cite{Brower:2005qw,Brower:2012vk}
\begin{equation}
  \label{eq:def_dw_eff}
  D^{4D}_\text{DW}(m)=\frac{1+m}{2}+\frac{1-m}{2}\gamma_5\text{sgn}(H_M).
\end{equation}
Here, $m$ is the quark mass, and the matrix sign function ``sgn'' is
approximated by 
\begin{equation}
  \label{eq:polar}
  \text{sgn}(H_M)=\frac{1-(T(H_M))^{L_s}}{1+( T(H_M))^{L_s}}
\end{equation}
with the transfer matrix $T(H_M)=(1-H_M)/(1+H_M)$.
The kernel operator $H_M$ is written as
\begin{equation}
  \label{eq:dw_kernel}
  H_M=\gamma_5\frac{\alpha D_W}{2+D_W},
\end{equation}
where $D_W$ is the Wilson-Dirac operator with a large negative mass
$-1/a$.
The scale parameter $\alpha$ is set to 2 in this work.
This corresponds to the M\"obius domain-wall fermion
\cite{Brower:2012vk}, while 
$\alpha=1$ gives the standard domain-wall fermion.
With this choice, the Ginsparg-Wilson relation is realized with a better precision
at a fixed $L_s$.
The sign function in (\ref{eq:polar}) is equivalent to the form
$\tanh(L_s\tanh^{-1}(H_M))$, 
which converges to the exact sign function in the limit 
$L_s\rightarrow \infty$. This is called the polar approximation.
In this limit, the Ginsparg-Wilson relation is exactly satisfied.
The details of our choice of the parameters are reported in \cite{Hashimoto:2014gta}.

The size of the violation of chiral symmetry for the M\"obius
domain-wall fermion may be quantified by the residual mass: 
\begin{equation}
  \label{eq:mres}
  m_{\rm res} =
  \frac{
    \langle\tr G^\dagger \Delta_\text{GW} G \rangle
  }{
    \langle\tr G^\dagger G \rangle
  },
\end{equation}
with
\begin{equation}
\Delta_\text{GW} \equiv 
\frac{\gamma_5}{2}\left[D^{4D}_\text{DW}(0)\gamma_5 + \gamma_5 D^{4D}_\text{DW}(0) - 2aD^{4D}_\text{DW}(0) \gamma_5 D^{4D}_\text{DW}(0)\right],
\label{eq:def_DeltaGW}
\end{equation}
where $G$ is the contact-term-subtracted quark propagator
\begin{equation}
  G = \frac{1}{1-m}
  \left( 
    (D^{4D}_\text{DW}(m))^{-1}-1
  \right).
\end{equation}
We confirm that the residual mass of the M\"obius domain-wall fermion
as defined in (\ref{eq:mres})
is roughly 5-10 times smaller than that of the standard domain-wall
Dirac operator at the same value of $L_s$ 
\cite{Hashimoto:2014gta}.

Even when the residual mass calculated as (\ref{eq:mres}) is small,
at a level of a few MeV or less,
the low-lying mode of $D_\text{DW}^\text{4D}$ may be
significantly affected by such small violation of the Ginsparg-Wilson relation
\cite{Cossu:2015kfa}. 
In fact, it was shown that the contribution to
the chiral condensate is in some cases dominated by the lattice
artifact that violates the Ginsparg-Wilson relation. 
Since we are interested in the details of the low-mode spectrum, we
need to carefully study such effects.
For that reason, we introduce the overlap fermion (with the same
kernel as the domain-wall fermion) and perform the reweighting to
eliminate the contamination from the lattice artifact.

One may improve the sign function approximation in
(\ref{eq:def_dw_eff}) by exactly treating the low-lying eigenmodes of
the kernel operator $H_M$, since the polar approximation is worse for
the low modes.
We compute $N_\text{th}$ lowest eigenmodes of the kernel operator $H_M$, 
and exactly calculate the sign function for this part of the spectrum.
Namely, we define
\begin{eqnarray}
  \label{eq:ov}
  D_\text{ov}(m) &=& 
  \sum_{|\lambda^M_i|<\lambda^M_\text{th}}
  \left[\frac{1+m}{2}+\frac{1-m}{2}\gamma_5\,\text{sgn}(\lambda^M_i)\right]
  |\lambda^M_i\rangle\langle\lambda^M_i|
\nonumber\\&&
  +
  D_\text{DW}^\text{4D}(m)
  \left[ 1 - 
    \sum_{\lambda^M_i<|\lambda^M_\text{th}|}
    |\lambda^M_i\rangle\langle\lambda^M_i|
  \right],
\end{eqnarray}
where $\lambda^M_i$ is the $i$-th eigenvalue of $H_M$ nearest to zero 
and $\lambda^M_\text{th}$ is a certain threshold.
We choose $\lambda^M_\text{th}$ = 400--600 MeV depending on the parameters.
With these choices, the violation of chiral symmetry 
is kept negligible, at the order of  $\sim$ 1~eV 
in our ensembles.

In this paper we slightly misuse the terminology and call
thus defined $D_\text{ov}$ the overlap-Dirac operator,
though the kernel is that of domain-wall fermion, {\it i.e.} the Shamir kernel.

Since the difference between $D_\text{DW}^\text{4D}$ and
$D_\text{ov}$ appears only in the treatment of the low modes of $H_M$,
we expect a good overlap in their relevant configuration spaces,
and a mild fluctuation of the reweighting factor between them.
This is indeed the case for the $16^3\times 8$ and $32^3\times 12$ lattices
we generated using $D_\text{DW}^\text{4D}$,
as we will see below.

\subsection{Configuration generation}
For the gauge part we use the tree-level improved Symanzik
gauge action \cite{Luscher:1985zq}.
We apply the stout smearing \cite{Morningstar:2003gk} three times 
on the gauge links with the $\rho$ parameter $\rho$ = 0.1
before computing the Dirac operators.
All the details on the choice of the parameters for these actions are
reported in our zero temperature studies~\cite{Cossu:2013ola, Kaneko:2013jla}.

\begin{table}[tbp]
  \begin{center}
    \begin{tabular}{c|c|c|c|c|c|c|c|c|c|c|c|c}
      $L^3\times L_t$ & $\beta$ & $ma$ & $L_s$&$m_\text{res}a$ & $T$ [MeV] & \#trj &$N_\text{conf}$ &
      $N^\text{eff}_\text{conf}$ &$N^{\text{eff}(2)}_\text{conf}$ 
      & $\tau^{\rm CG}_\text{int}$ & $\tau^{\rm top}_\text{int}$& $M_{PS}L$ \\ \hline\hline
      $16^3\times 8$ &$4.07$& 0.01    &$12$& 0.00166(15) & 203(1) &6600& 239 &11(13)&45(8)  &70 &25(6)&5.4(3)\\
      $16^3\times 8$ &$4.07$& 0.001  &$24$& 0.00097(43) & 203(1) &12000& 197 & 7 (7) &14(3) &315 &23(4)& 5.3(4)\\

      $16^3\times 8$ &$4.10$& 0.01    &$12$& 0.00079(5) & 217(1) &7000& 203 &23(7)&150(17) &134 &30(10)& 6.9(5)\\
      $16^3\times 8$ &$4.10$& 0.001  &$24$& 0.00048(14) & 217(1) &12000& 214 &31(10)&121(10)&104 &24(4)& 6.3(9)\\      \hline

      $32^3\times 8$ &$4.07$& 0.001   &$24$& 0.00085(9) & 203(1) &4200& 210 & 10(3)$^*$&--&128 &18(4)&11.7(9)\\

      $32^3\times 8$ &$4.10$& 0.01    &$12$& 0.0009(5) & 217(1) &3800& 189 & 9(4)$^*$&--&125&30(10)& 12.6(5)\\
      $32^3\times 8$ &$4.10$& 0.005   &$24$& 0.00053(4) & 217(1) &3100& 146 &20(4)$^*$&-- &84&24(9)& 11.6(7)\\
      $32^3\times 8$ &$4.10$& 0.001   &$24$& 0.00048(5) & 217(1) &7700& 229 &18(5)$^*$&-- &10&23(5)& 12.3(9)\\      \hline

      $32^3\times 12$ &$4.18$& 0.01   &$16$& 0.00022(5) & 172(1) &2600&  (319) &--&--&-- &--& 5.8(1)\\

      $32^3\times 12$ &$4.20$& 0.01   &$16$& 0.00020(1) & 179(1) &3400&  (341) &--&--&-- &--& --\\

      $32^3\times 12$ &$4.22$& 0.01   &$16$& 0.00010(1) & 187(1) &7000&  (703) &--&--&-- &--& 5.4(2)\\

      $32^3\times 12$ &$4.23$& 0.01   &$16$& 0.00008(1) & 191(1) &5600& 51 &28(4)&38(5)&240 &120(50)&-- \\
      $32^3\times 12$ &$4.23$& 0.005   &$16$& 0.00012(1) & 191(1) &10300& 206 &22(2)&27(2)&131 &160(140)&-- \\
      $32^3\times 12$ &$4.23$& 0.0025  &$16$& 0.00016(4) & 191(1) &9400& 195 &16(2)&255(31)&85 &110(30)&-- \\

      $32^3\times 12$ &$4.24$& 0.01   &$16$& 0.00008(1) & 195(1) &7600& 49 &23(5)& 36(5)&125 &100(40)&6.8(5)  \\
      $32^3\times 12$ &$4.24$& 0.005   &$16$& 0.00010(2) & 195(1) &9700& 190 &9(18)&53(6)&84 &130(30)&-- \\
      $32^3\times 12$ &$4.24$& 0.0025  &$16$& 0.00011(2) & 195(1) &16000& 188 &8(10)&7(1)&618 &80(20)& 6(2)\\

      \hline
    \end{tabular}
    \caption{
      Summary of simulated ensembles.
      The residual mass $m_\text{res}a$ is calculated using the definition (\ref{eq:mres}).
      \#trj denotes the number of trajectories.
      $N_\text{conf}$ presents the number of configurations generated
      (those with parenthesis are not used in the main analysis of this work 
      but used for the estimate of the critical temperature).
      $N^\text{eff}_\text{conf}$ and $N^{\text{eff}(2)}_\text{conf}$ are the effective statistics
      after the overlap/domain-wall reweighting, which are defined by (\ref{eq:Neff}) and (\ref{eq:Neff2})
      (data with $^*$ are measured by low-mode approximation of the reweighting).
      $\tau^{\rm CG}_\text{int}$ and $\tau^{\rm top}_\text{int}$ are the integrated 
      auto-correlation time of the CG iteration count and topological charge,
      respectively, in the units of the molecular dynamics time.
      $M_{PS}L$ is the screening mass of the pseudoscalar correlator, multiplied by the lattice size $L$.
      \label{tab:setup}
    }
  \end{center}
\end{table}

\begin{figure}[tbp]
  \centering
  \includegraphics[width=8cm]{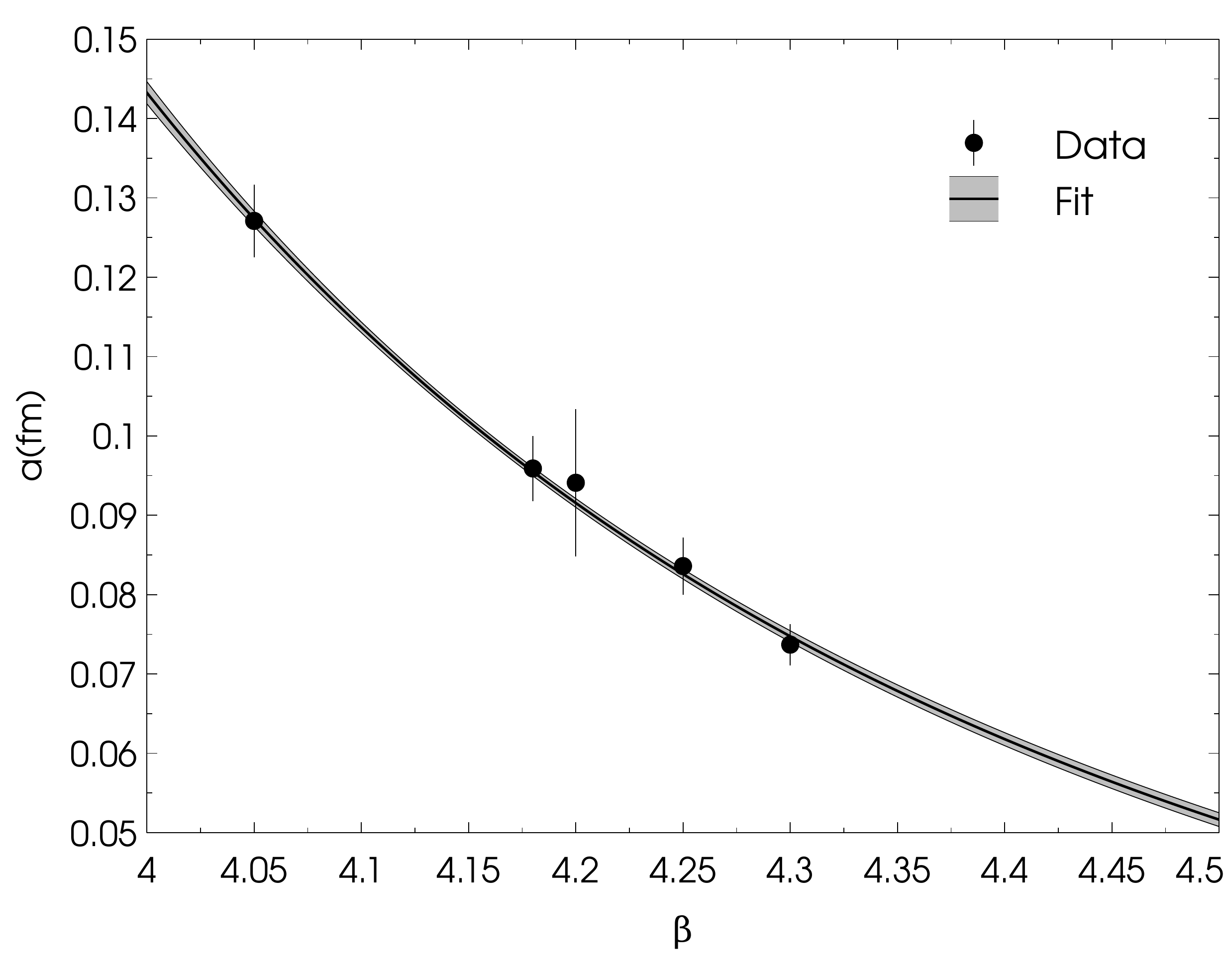}
  \caption{
    Lattice spacing $a$ as a function of $\beta$, for the
    choice of lattice action in this work, estimated using the Wilson flow scale $t_0$.
    The data are interpolated by a perturbative
    prediction given in (\ref{eq:betadep}).
    The band represents the error in the estimate, which is
    negligible ($<0.5\%$) for the ensembles discussed in this
    work.
  }
  \label{fig:latsp}
\end{figure}

Our simulation setup is summarized in Table~\ref{tab:setup}. 
The lattice spacing $a$ is estimated by the Wilson flow
on a few selected zero-temperature configurations.
We use the reference flow time $t_0=(0.1539\;\text{fm})^2$ determined by 
the ALPHA collaboration \cite{Sommer:2014mea} for $N_f=2$ QCD.
The results at different quark masses are extrapolated to the chiral limit 
and the data are interpolated in $\beta$ assuming perturbative running
 \cite{Edwards:1997xf, Larin:1993tp, Caswell:1974gg, Jones:1974mm, Egorian:1978zx}:
\begin{eqnarray}
\label{eq:betadep}
a &=& c_0 f(g^2)(1+c_2\hat{a}(g)^2),\;\;\; \hat{a}(g)^2\equiv [f(g^2)]^2,\nonumber\\
f(g^2)&\equiv& (b_0g^2)^{-b_1/2b_0^2}\exp\left(-\frac{1}{2b_0 g^2}\right),
\nonumber\\
b_0&=&\frac{1}{(4\pi)^2}\left(11-\frac{2}{3}N_f\right),\;\;\;
b_1=\frac{1}{(4\pi)^4}\left(102-\frac{38N_f}{3}\right),
\end{eqnarray}
where $g^2=6/\beta$, $N_f=2$, and $c_0$ and $c_2$ are free parameters of the fit.
The result is plotted in Fig.~\ref{fig:latsp} with 
our estimates $c_0=6.9(2)$ and $c_2=6.1(6)\times 10^3$.
The simulated lattice spacing covers the range between 0.074~fm ($\beta=4.30$) and 0.127~fm ($\beta=4.05$).
Our estimates of the temperature on $L_s$ = 8 and 12 lattices are 
listed in Table~\ref{tab:setup}.

\begin{figure}[tbp]
  \centering
  \includegraphics[width=8cm]{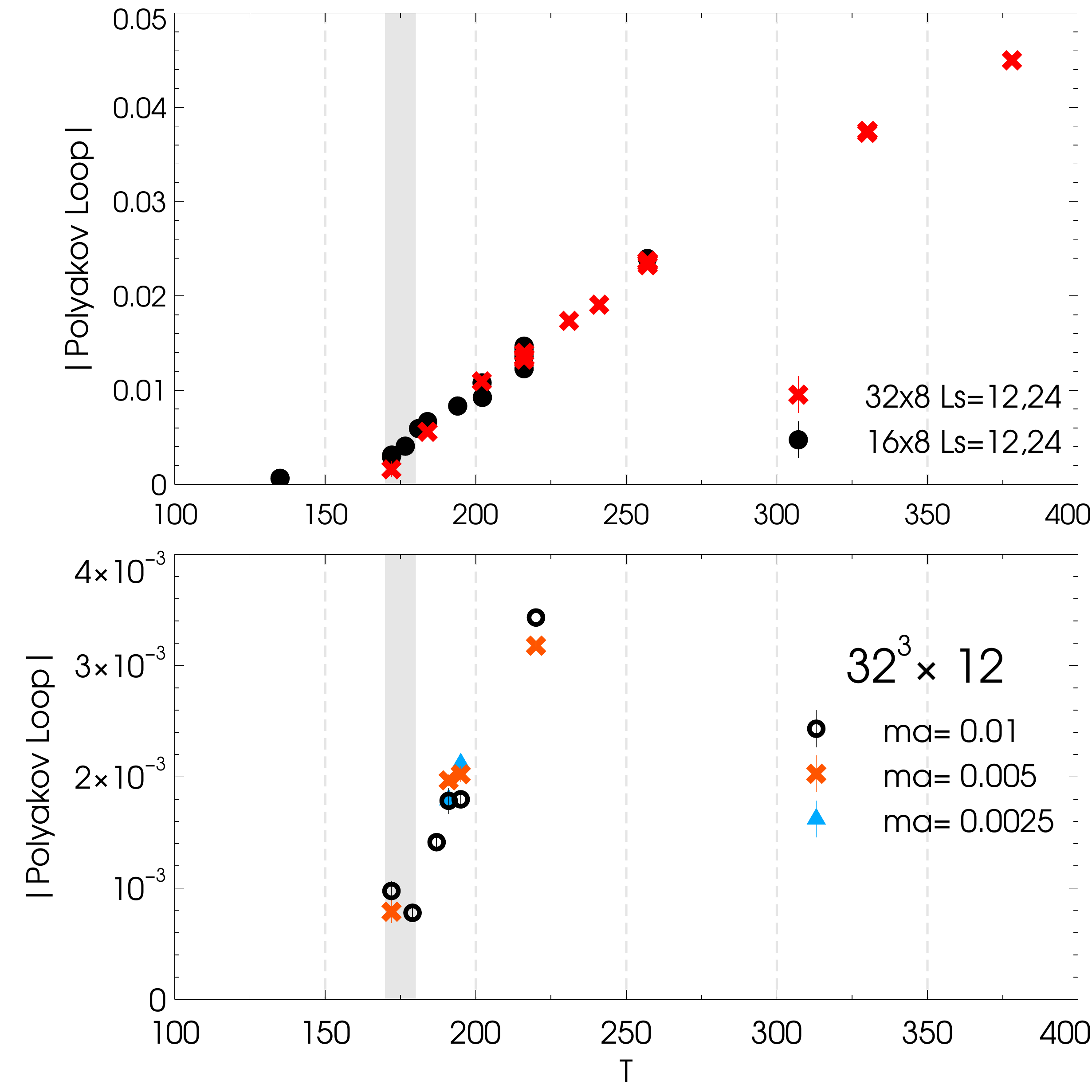}
  \caption{
    The Polyakov loop expectation value as a function of temperature.
    The results for $L_t=8$ (top panel) and those for $L_t=12$ (bottom) are shown.
    The critical temperature $T_c$ is estimated as 175(5) MeV,
    which is shown by a shadow band.
  }
  \label{fig:PL}
\end{figure}

We estimate the critical temperature as $T_c=175(5)$ MeV from
the inflection point of the Polyakov loop,
as shown in Fig.~\ref{fig:PL}.
We confirm that the ensembles listed in Table~\ref{tab:setup} are at
or above the chiral phase transition.

\begin{figure*}[tbp]
  \centering
  \includegraphics[width=8cm]{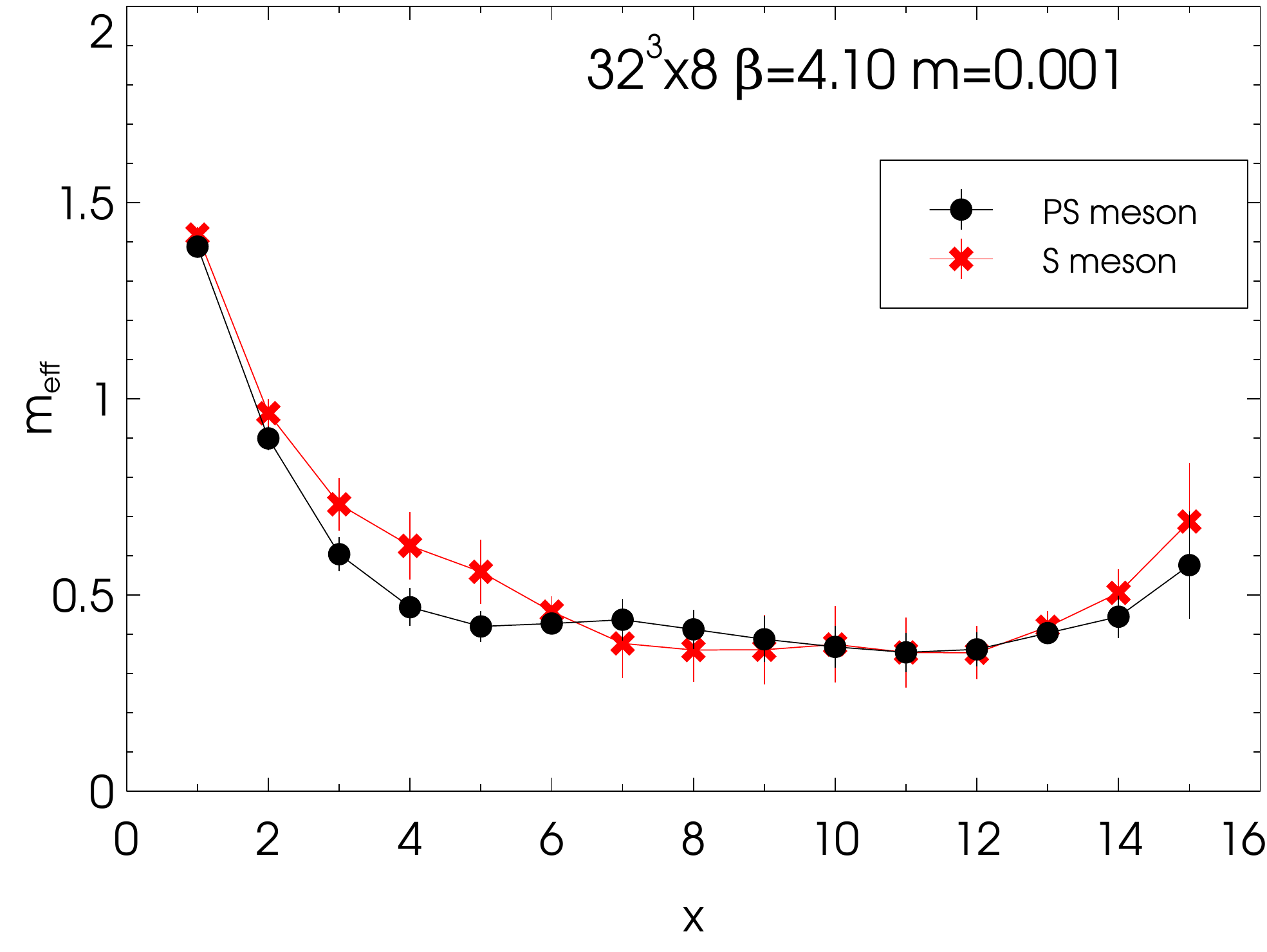}
 \includegraphics[width=8cm]{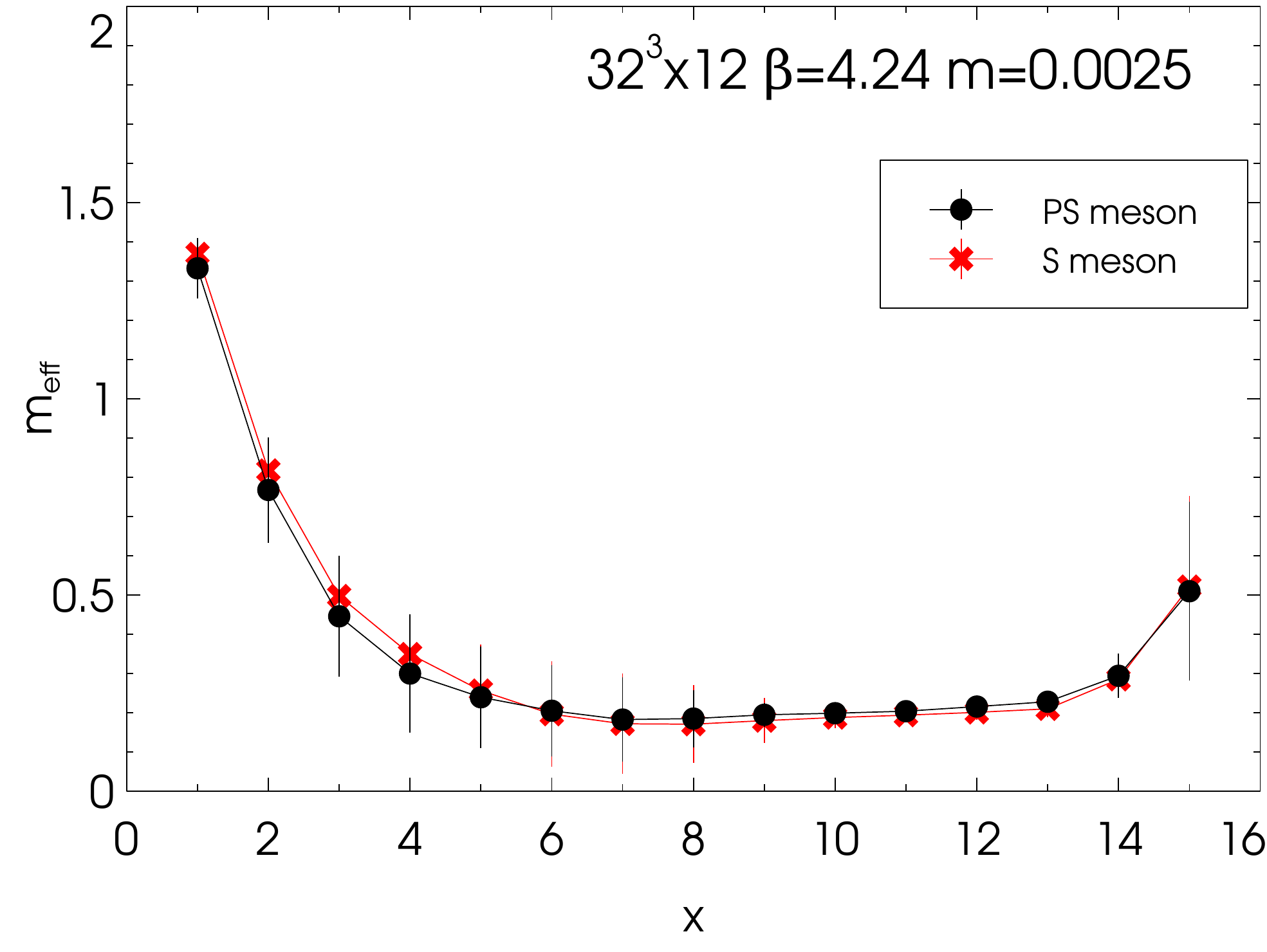}
  \caption{
    Effective mass plots of the spatial correlator 
    at $\beta =4.10$, $m=0.001$ on the $32^3\times 8$ 
    lattice (left panel),
    and those at $\beta =4.24$, $m=0.0025$ on the $32^3 \times 12$ lattice (right).
    The correlation functions  in the pseudoscalar (black) and scalar (red) channels are shown.
  }
  \label{fig:screeningmass}
\end{figure*}

The physical lattice size $L$ is 2--4~fm in the spatial directions.
We confirm by the calculation of the pseudoscalar correlators
that the correlation length in the spatial direction $1/M_{PS}$ 
is sufficiently small 
compared to the spatial lattice size $L$.
The values of $M_{PS}L$ are listed in Table~\ref{tab:setup}.
We also present some effective mass plots in Fig.~\ref{fig:screeningmass}.
Due to a small correlation length of the finite temperature system,
no significant finite volume effects are expected even 
for those ensembles at smallest quark masses.

The bare quark mass is chosen in the range from 2 to 25~MeV.
The residual mass (\ref{eq:mres}) in our simulations is 
$\leq$ 1~MeV around $T_c$ and even smaller at higher temperature.
We have data at two values of $L_s$ = 8 and 12 to check the
discretization effects.

The ensembles are generated with the standard Hybrid Monte Carlo method.
We estimate the autocorrelation time   
using the CG counts in the inversion of the fermion matrix.
Depending on ensembles, it is around 70-600 in the unit of the
molecular dynamics time.
Statistical error is estimated using the jackknife method
with the bin size greater than 
the integrated auto-correlation time of the observable.
The number of measurement is $O(100)$ depending on the 
ensemble as listed in the column of \#conf.
Each measurement is separated by 20--100 molecular dynamics time.

\begin{figure}[tbp]
  \centering
  \includegraphics[width=8cm]{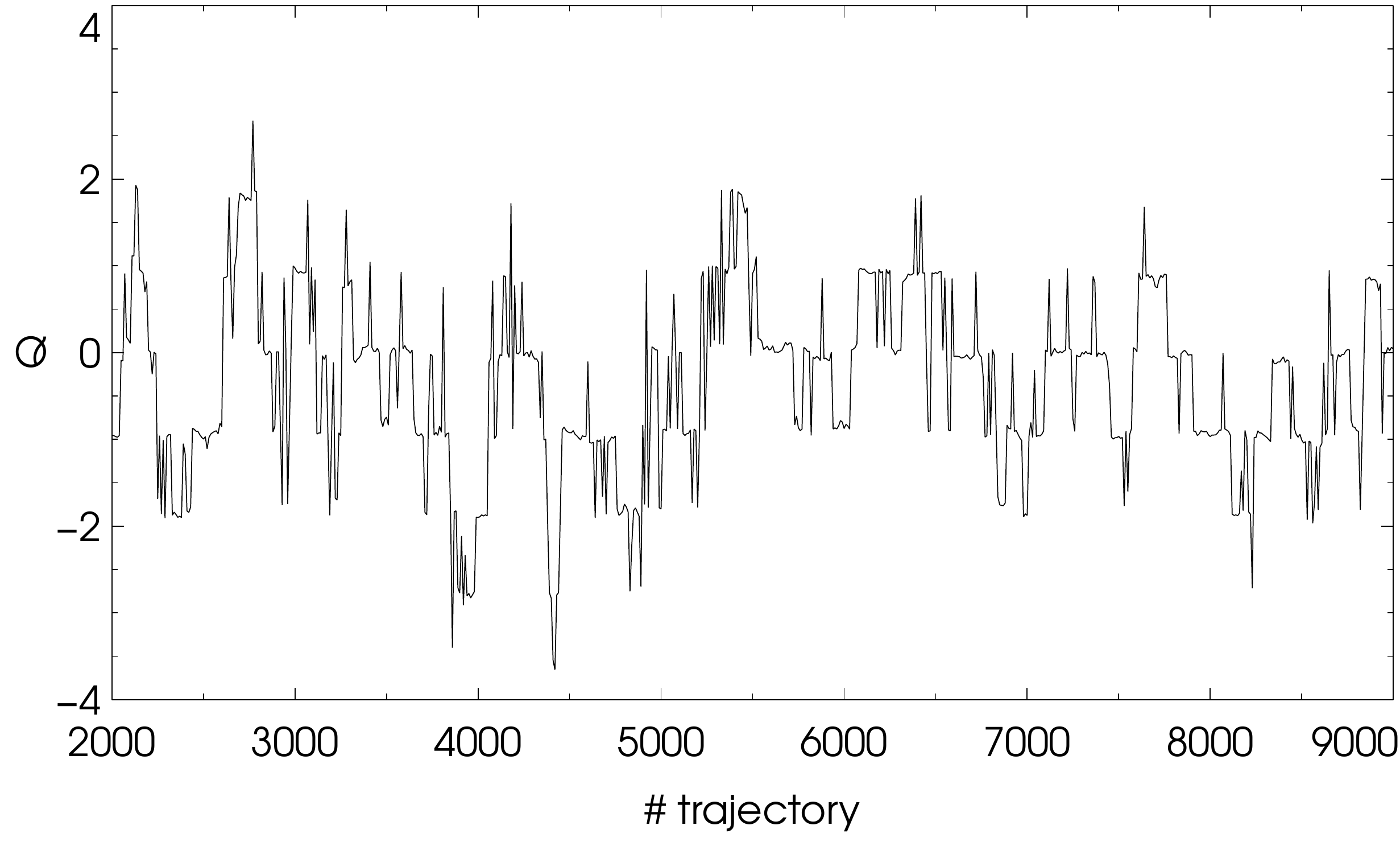}
  \caption{
    History of the topological charge for $L^3\times L_t=32^3\times 12$,
    $\beta=4.24$, $m=0.01$.
  }
  \label{fig:top_hist}
\end{figure}

In order to check whether the topology tunneling frequently occurs in our simulations, 
we monitor the topological charge $Q$ of each configuration.
Here we use the field theoretical definition for $Q$ 
\begin{equation}
Q = \frac{1}{32\pi^2}\sum_x \epsilon^{\mu\nu\rho\sigma}{\rm Tr} F_{\mu\nu}(x)F_{\rho\sigma}(x),
\end{equation}
where the field strength $F_{\mu\nu}$ is defined using the clover leaf construction,
measured after the Wilson flow \cite{Luscher:2010iy} of its flow time $ta^2\simeq$ 5
(the measurement of $Q$ is stable for $ta^2 \gtrsim 3$).
We confirm that it changes frequently along the simulations.
One example is shown in Figure~\ref{fig:top_hist}.
The autocorrelation time for the topological charge is 
also listed in Table~\ref{tab:setup}.

For the configurations generated, we compute the low-lying
eigenvalues of the 4D hermitian effective operators
$H_\text{DW}^\text{4D}(m)\equiv\gamma_5 D_\text{DW}^\text{4D}(m)$ and
$H_\text{ov}(m)\equiv\gamma_5 D_\text{ov}(m)$
using the implicitly restarted Lanczos algorithm with $O(100)$ Krylov vectors\footnote{
  In our typical simulations, $H_\text{ov}(m)$ eigenvalues/eigenvectors
  computation fore one configuration costs roughly the same as 10-30 trajectories of the HMC run.
  That for $H_M$ is 10 times faster.
  However, these numerical costs strongly depend on temperature and how many eigenvalues we need.
}.
From the eigenvalues of $H_\text{ov}(m)$, 
we can also extract the number of chiral zero-modes 
$n_+$ with positive chirality and $n_-$ with negative chirality.
For each zero mode of $D_\text{ov}(0)$ with $\pm$ chirality, 
we have a mode whose eigenvalue of $H_\text{ov}(m)$ is $\pm m$.
Since these zero modes are generally isolated, while
other non-zero modes make $\pm \lambda^{(m)}$ pairs, 
our numerical determination of $n_+$ and $n_-$ is quite robust,
and thus we can determine the topological index $\nu=n_+-n_-$
of the overlap Dirac operator, as well as the number of
the zero modes $N_0=n_+ + n_-$.

\subsection{Overlap/domain-wall reweighting}\label{sec:Reweighting}

The expectation value of an observable $\mathcal{O}$ 
with the dynamical overlap fermion can be  estimated  by the reweighting as
\begin{equation}
  \langle \mathcal{O}\rangle_\text{ov}
  = \frac{\langle \mathcal{O}R\rangle_{\rm DW}}{\langle R\rangle_{\rm DW}},
\end{equation}
where $\langle\cdots\rangle_\text{DW}$ and
$\langle\cdots\rangle_\text{ov}$ 
denotes the ensemble average with the \mobius domain-wall and overlap sea
quarks, and $R$ is the reweighting factor
\begin{equation}
  \label{eq:R}
  R \equiv 
  \frac{\det[H_\text{ov}(m)]^2}{\det[H_\text{DW}^\text{4D}(m)]^2}
  \times
  \frac{\det[H_\text{DW}^\text{4D}(1/4a)]^2}{\det[H_\text{ov}(1/4a)]^2}.
\end{equation}
The second factor 
$\det[H_\text{DW}^\text{4D}(1/4a)]^2/\det[H_\text{ov}(1/4a)]^2$
in (\ref{eq:R}) is introduced to cancel the noise from high modes at
the cutoff scale \cite{Fukaya:2013vka}.
It corresponds to adding fermions and ghosts of
a cutoff scale mass $1/4a$, and therefore
does not affect the low-energy physics we are interested in.
The reweighting factor is stochastically 
estimated~\cite{Hasenfratz:2008fg} with Gaussian noise fields $\xi_i$ and  $\xi_i'$,
\begin{equation}
  R =\frac{1}{N}\sum_{i=1}^N \exp\left[
    -\xi^\dagger_i [H_\text{DW}^\text{4D}(m)]^2[H_\text{ov}(m)]^{-2}\xi_i
    -\xi'^\dagger_i [H_\text{DW}^\text{4D}(1/2a)]^{-2}[H_\text{ov}(1/2a)]^{2}\xi'_i
  \right],
\end{equation}
with a few noise samples for each configuration.

\begin{figure}[tbp]
  \centering
  \includegraphics[width=10cm]{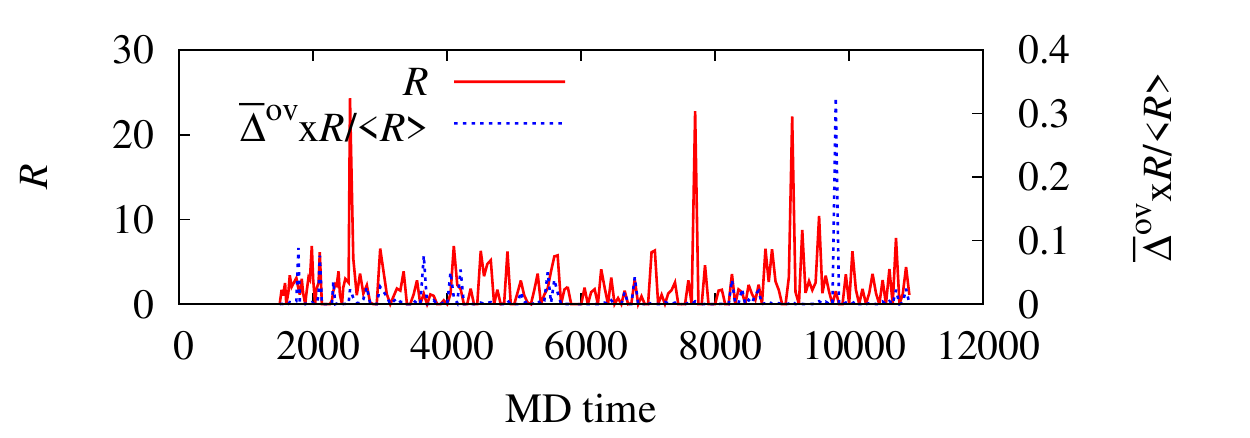}
\includegraphics[width=10cm]{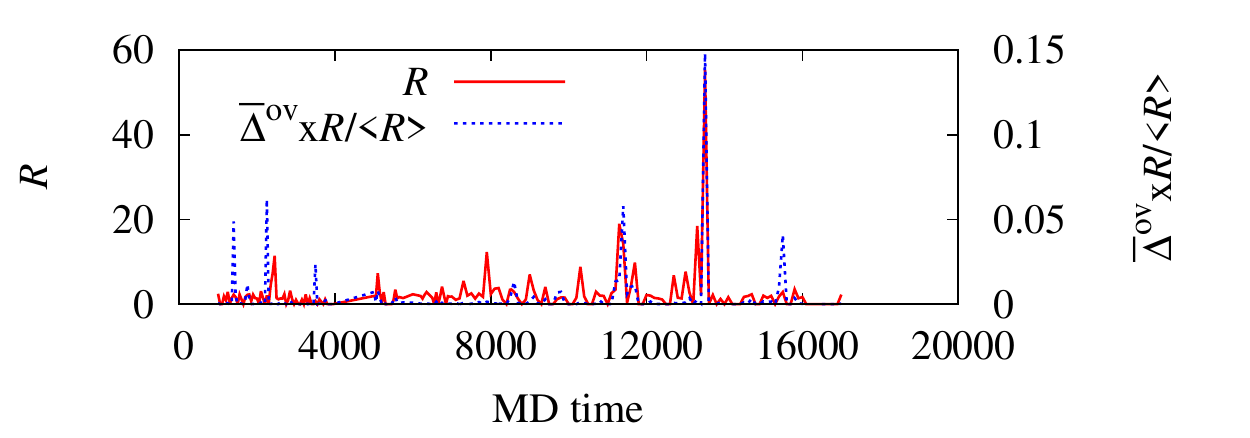}
  \caption{
    History of the reweighting factor $R$ (solid) and that of $\bar{\Delta}_{\pi-\delta}^{\rm ov}\times R/\langle R\rangle$ (dashed) for 
    $L^3\times L_t=32^3\times 12$ ensembles at $\beta=4.23$(top), 4.24(bottom) with the same bare quark mass $m=0.0025$.
    The definition of $\bar{\Delta}_{\pi-\delta}^{\rm ov}$ is given by Eq.~(\ref{eq:Deltabar}).
  }
  \label{fig:reweighting_factor}.
\end{figure}

The reweighting is effective when the factor $R$ does not
fluctuate too much.
Since the factor scales exponentially as a function of the volume 
of the lattice, the relevant matrix
$[H_\text{DW}^\text{4D}(m)]^2[H_\text{ov}(m)]^{-2}$ 
needs to be close to an identity operator.
Our operator $D_\text{ov}(m)$ is designed to satisfy this condition,
{\it i.e.} only the treatment of the near-zero eigenmodes of the
kernel operator is different.
It is however not known how such difference affects $R$
until we actually compute it.
Figure~\ref{fig:reweighting_factor} shows examples of the Monte
Carlo history of $R$.
It turns out that the maximum of $R$ is at the level of $O(10)$ on 
$16^3\times 8$ and $32^3\times 12$ lattices,
which does not destroy the ensemble average when we have $O(100)$ samples. 
To assess the quality of the reweighting 
we define the effective  number of configurations \cite{Ogawa:2005jn} by
\begin{eqnarray}
\label{eq:Neff}
N^\text{eff}_\text{conf} = \frac{\langle R\rangle}{R_\text{max}},
\end{eqnarray}
where $R_\text{max}$ is the maximum value of the reweighting factor
in the ensemble. However, as shown in the same plot in Fig.~\ref{fig:reweighting_factor} 
it turns out that $R_\text{max}$ does not necessarily 
coincide with the peak of the observable $\mathcal{O}R$, {\it e.g.} 
$\mathcal{O}=\bar{\Delta}_{\pi-\delta}^{\rm ov}$ as  defined later.
Therefore, we also measure 
\begin{eqnarray}
\label{eq:Neff2}
N^{\text{eff}(2)}_\text{conf} = \frac{\langle R\rangle}{R'_\text{max}},
\end{eqnarray}
with $R'_\text{max}$ the reweighting factor which gives the maximum value of 
$\bar{\Delta}_{\pi-\delta}^{\rm ov}\times R$ in the ensemble.
Both $N^\text{eff}_\text{conf}$ and $N^{\text{eff}(2)}_\text{conf}$ are listed in Table~\ref{tab:setup}.
$N^{\text{eff}(2)}_\text{conf}$ is larger than $N^\text{eff}_\text{conf}$ except for
the configurations at $\beta=4.24$ and $m=0.0025$.

In particular, on the $16^3\times 8$ lattices, the reweighting factors
are stable enough that we can choose different quark masses from that of the original ensemble: $m=0.005$ 
on $m=0.01$ \mobius domain-wall ensembles.

There are some configurations for which the reweighting factor
is essentially zero, say $R<10^{-3}$.
For these configurations, we find chiral zero-modes for the
overlap-Dirac operator.
They are suppressed as the fermion determinant contains a factor
$(am)^2$ from the zero-mode and the next lowest eigenvalues are also
smaller compared to the corresponding eigenvalues of the M\"obius
domain-wall Dirac operator.
We note that the pairing of the positive and negative eigenvalues of
$H_\text{ov}$ is precisely satisfied other than the exact zero modes.
With $H_\text{DW}^\text{4D}$ such correspondence is hardly visible
especially for the coarser lattices at $L_t$ = 8.

For the large-volume lattices of size $32^3\times 8$, we found that
the reweighting as described above are not effective.
On these lattices, the lattice spacing is relatively large, $a\simeq$
0.11~fm, and the difference between $D_\text{DW}^\text{4D}$ and
$D_\text{ov}$ is more significant.
With larger spatial volume, such difference is enhanced exponentially. 
For these lattices, we introduce a reweighting scheme that takes only
the low-mode part of the eigenvalue spectrum.
Namely we approximate the reweighting factor by
\begin{equation}
  \label{eq:lowmode_reweighting}
  R_\text{low} 
  =  \frac{
    \prod_{k=1}^{N_{th}} (\lambda_{\text{ov},k}^{(m)})^2
  }{
    \prod_{k=1}^{N_{th}} (\lambda_{\text{DW},k}^{(m)})^2
  },
\end{equation}
where $\lambda_{\text{ov},k}^{(m)}$ and $\lambda_{\text{DW},k}^{(m)}$ denote the
$k$-th lowest eigenvalue of the hermitian operators
$H_\text{ov}(m)$ and $H_\text{DW}^\text{4D}(m)$, respectively.
The number of the eigenvalues included $N_{th}$ is 40 
except for $N_{th}=10$ at $\beta=4.10$ and $m=0.01$.
The threshold is around $0.1/a \sim 160$ MeV.

The low-mode reweighting (\ref{eq:lowmode_reweighting}) corresponds to introducing 
an extra ultraviolet modification to the fermion determinant ratio (\ref{eq:R}).
However, its modification does not distort
the continuum limit since $D_\text{DW}^\text{4D}$ precisely converges
to $D_\text{ov}$ in that limit as the probability of having 
low-lying modes of $2H_M$ less than $\lambda^M_\text{th}$ vanishes.
Namely, both of $R$ and $R_\text{low}$ are guaranteed to converge to unity in the continuum limit,
and share the same continuum limit.
We confirm on the small $16^3\times 8$ lattice where the full reweighting
is available, that $R$ and $R_\text{low}$ give consistent results for the Dirac spectrum.
We also confirm that our observable for the \ua symmetry breaking
is dominated by the lowest modes much below the threshold of $0.1/a$,
as will be discussed in Sec.~\ref{sec:sus}.

As will be shown below, our target of this work, $U(1)_A$ sensitive quantities, are
sensitive to the overlap/Domain-wall reweighting.
However, the reweighting does not affect those insensitive to the $U(1)_A$ symmetry.
For example, we find that the plaquette changes only by less than $0.3$\%,
which is much smaller than its statistical error (of reweighted plaquettes).
Table. \ref{tab:polyakov_comparison_rew} is a comparison of the Polyakov loop with and without the reweighting.
This table shows that 
for $U(1)_A$ insensitive quantities, the overlap fermions and
\mobius domain-wall fermions are essentially the same.
\begin{table}[htb]
\center
    \begin{tabular}{c|c||c|c}
$\beta$ & $am_\text{ud}$ & 
$\av{L}$  & 
$\av{L}_\text{rew}$  
\\  \hline \hline
4.07 & 0.01 & 
0.01032 (45) &
0.01023 (18)  
\\ 
4.07 & 0.001 & 
0.01147 (27) &
0.0117 (15) 
\\ \hline
4.10 & 0.01 & 
0.01457 (34) & 
0.0141 (13) 
\\ 
4.10 & 0.001 & 
0.01294 (45) &
0.0130 (11) 
 \\  \hline
4.23 & 0.01 & 
0.00225 (16) &
0.00254 (54)   
 \\ 
4.23 & 0.005 & 
0.00495 (75) &
0.00435  (92)   
 \\ 
4.23 & 0.0025 & 
0.00262 (23) &
0.00235 (57)   
 \\  \hline
4.24 & 0.01 & 
0.00233 (18) &
0.00245 (77) 
 \\ 
4.24 & 0.005 & 
0.00788 (61)& 
0.0076 (13) 
 \\ 
4.24 & 0.0025 & 
0.00400 (48) &
0.00367 (75) 
 \\  \hline
    \end{tabular}
\caption{Polyakov loop for original configurations $\av{L}$
and reweighted one $\av{L}_\text{rew}$.
Here coarser lattice data is for $L^3=16^3$.
\label{tab:polyakov_comparison_rew} }
\end{table}


\section{Dirac spectrum}
\label{sec:Dirac}
In this section, we study the Dirac spectrum $\rho(\lambda)$, 
which is tightly related to both of the \su and \ua symmetries\footnote{
For our recent study of the chiral symmetry breaking at zero temperature, see Ref.~\cite{Cossu:2016eqs}.
}.
We compute the eigenvalues $\lambda_k^{(m)}$ of the massive
operators $H_\text{DW}^\text{4D}(m)$ and $H_\text{ov}(m)$, 
and evaluate those of the massless operators using
\begin{equation}
  \lambda_k =
  \frac{\sqrt{(\lambda_k^{(m)})^2-m^2}}{\sqrt{1-m^2}}.
  \label{eq:massless_eigenvalue}
\end{equation}
When the Ginsparg-Wilson relation is satisfied,
$\lambda_k$ is exactly the same as the corresponding 
eigenvalue of the massless Dirac operator. 
We apply the same formula to the M\"obius domain-wall Dirac eigenvalues, though
the Ginsparg-Wilson relation is not exact.
We confirm that $|\lambda_k^{(m)}|>m$ is always satisfied
and the effect of $m_\text{res}$ is invisible with our resolution of the
Dirac eigenvalue density explained below.

\begin{figure*}[tbp]
    \centering
    \includegraphics[width=8.0cm]{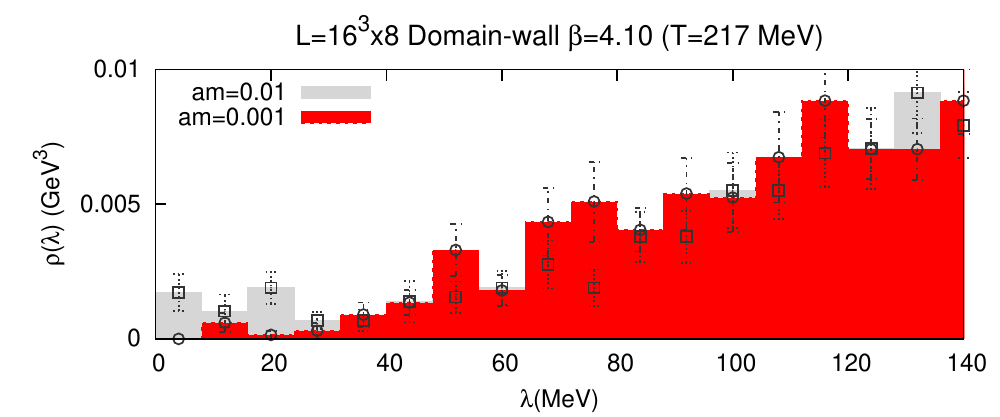}
    \includegraphics[width=8.0cm]{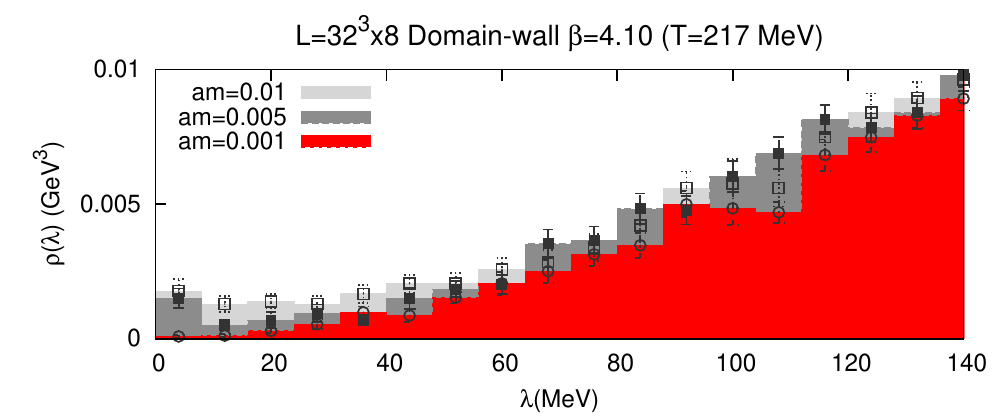}
    \includegraphics[width=8.0cm]{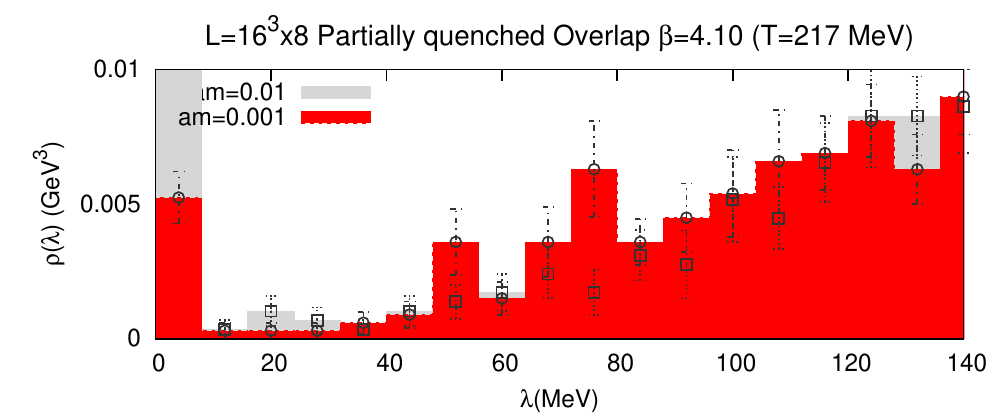}
    \includegraphics[width=8.0cm]{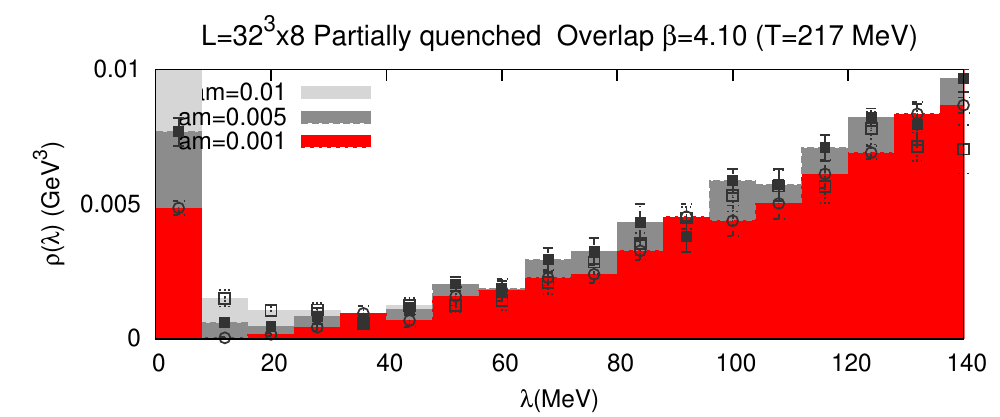}
    \includegraphics[width=8.0cm]{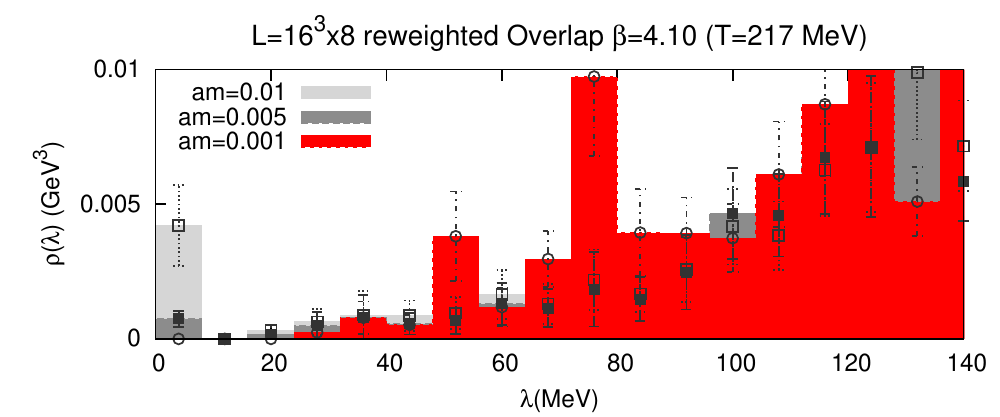}
    \includegraphics[width=8.0cm]{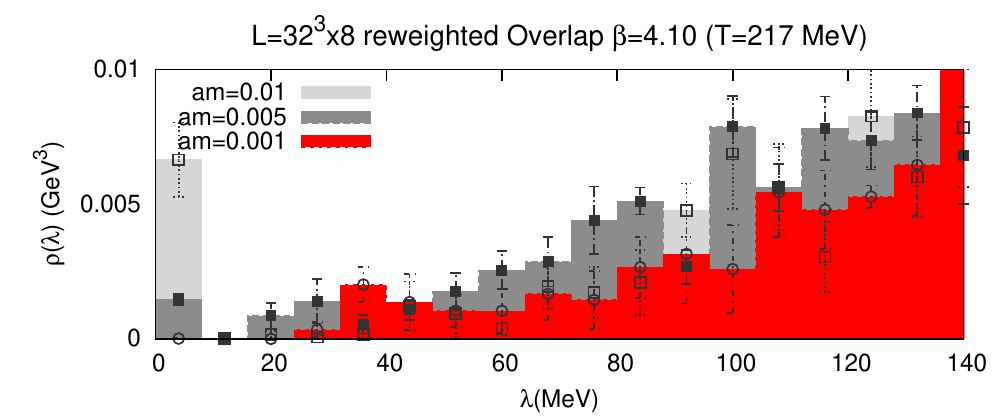}
  \caption{
    Eigenvalue spectrum of the M\"obius domain-wall (top panels),
    partially quenched overlap with M\"obius domain-wall sea quarks (middle)
    and the (reweighted) overlap (bottom) Dirac operators. 
    Data at $\beta$ = 4.10 ($T\sim$ 217~MeV) on $16^3\times 8$ (left panels)
    and $32^3\times 8$ (right) are shown.
    The data for $m=0.005$ on the $16^3\times 8$ lattice are obtained by reweighting 
    on the $m=0.01$ ensemble.
  }
  \label{spectrum:b4.10}
\end{figure*}

Figure~\ref{spectrum:b4.10} shows the eigenvalue histograms
of the M\"obius domain-wall (top panels),
partially quenched overlap with M\"obius domain-wall sea quarks (middle)
and (reweighted) overlap (bottom) Dirac operators. 
Data at $\beta$ = 4.10 ($T\sim$ 217~MeV) on the $16^3\times 8$ lattice 
are shown on the left panels, 
and those on the $32^3\times 8$ lattice are shown on the right panels.
Here we count the number of eigenvalues in a bin
$[\lambda-\mbox{4~MeV},\lambda+\mbox{4~MeV}]$ and rescale them by $1/V$ to obtain
the eigenvalue density $\rho(\lambda)$ in the physical unit.
When the data for different sea quark masses are plotted together,
the heavier mass data are shown by shaded histograms.
When there are exactly chiral zero-modes, they are included in the lowest bin.

The M\"obius domain-wall Dirac operator spectrum shows 
a mild slope towards zero at the lightest quark masses near the chiral limit.
This slope is consistent with $\lambda^3$, which was also 
reported in \cite{Chiu:2013wwa} employing the optimal domain-wall fermions.
The reweighted overlap Dirac operator histograms look similar but
we can see a stronger suppression of the near zero modes: the 
first three bins are consistent with zero,
which is consistent with our previous work \cite{Cossu:2013uua}.

In contrast to a qualitative agreement of the \mobius domain-wall and overlap 
Dirac operators, a striking difference is seen in 
the data for the partially quenched overlap or the overlap Dirac 
spectrum without reweighting: a sharp peak is found at the lowest bin,
which does not disappear even at the lightest quark mass.
The appearance of such a peak structure, mainly coming from 
the chiral zero modes, 
is known from previous works 
with overlap fermions on pure gauge configurations 
(see {\it e.g.} \cite{Cossu:2013uua}). 
Recently, a similar structure was reported in the overlap Dirac spectrum
on the ensembles generated by the HISQ action \cite{Dick:2015twa}.
Since such a peak does not appear in the \mobius domain-wall 
and reweighted overlap eigenvalues, 
they are likely an artifact of
partial quenching.

\begin{figure*}[tbp]
    \centering
    \includegraphics[width=8.0cm]{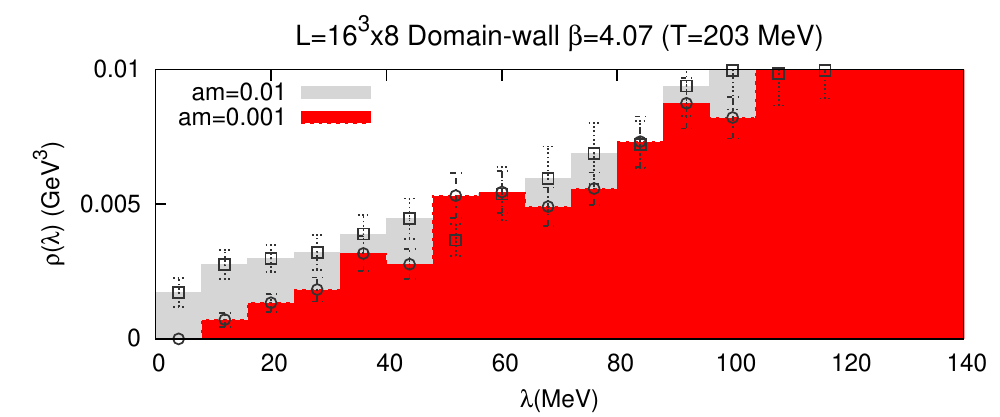}
    \includegraphics[width=8.0cm]{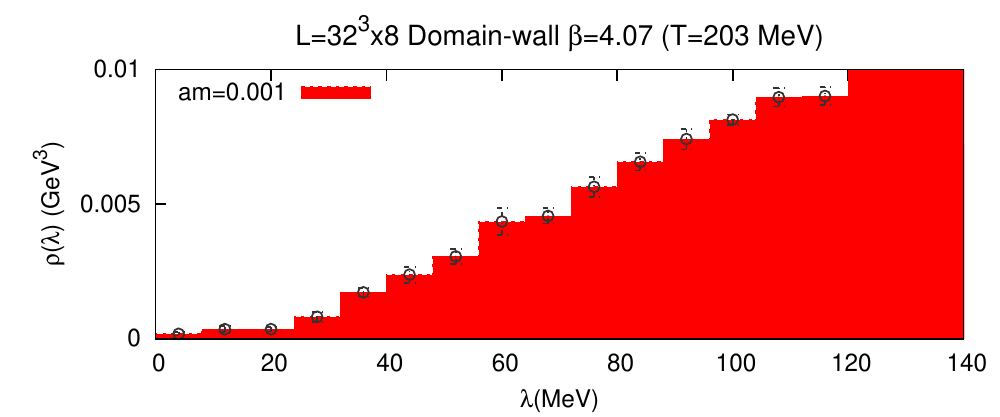}
    \includegraphics[width=8.0cm]{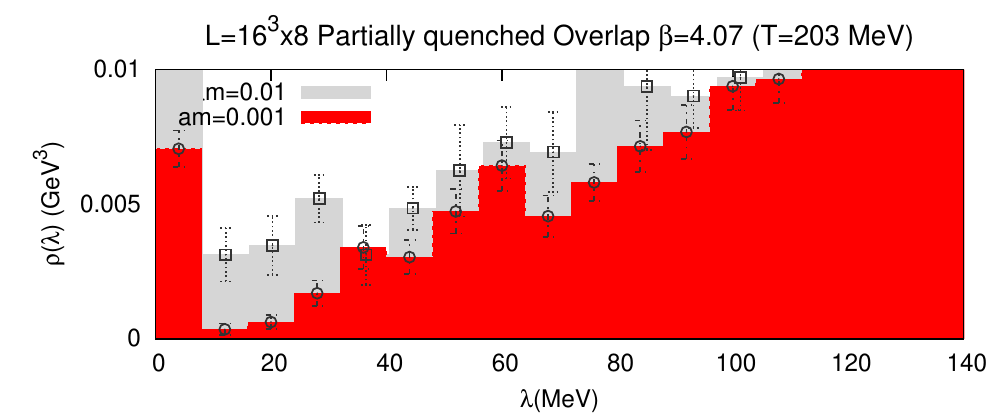}
    \includegraphics[width=8.0cm]{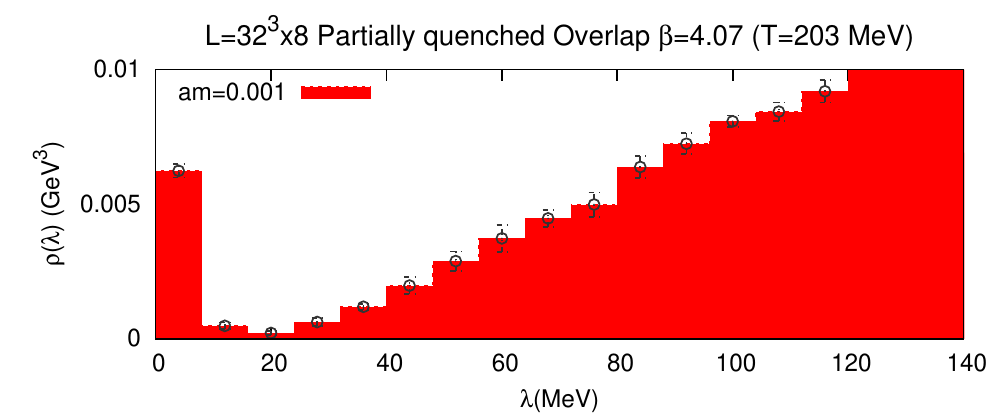}
    \includegraphics[width=8.0cm]{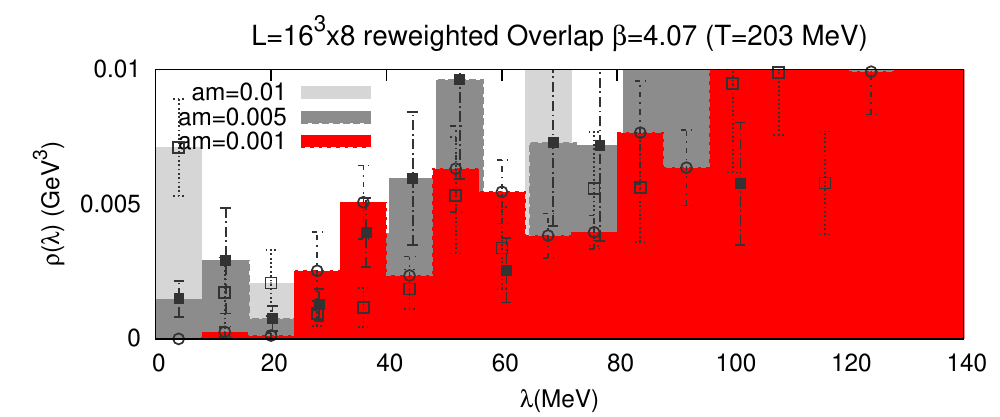}
    \includegraphics[width=8.0cm]{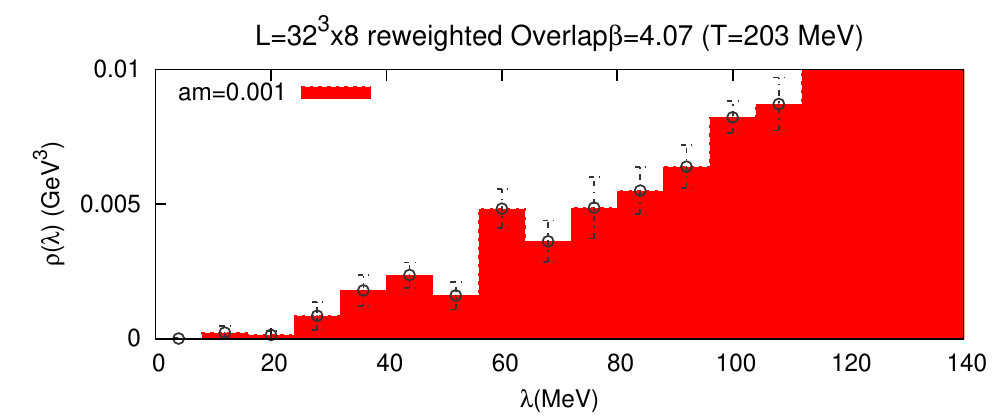}
  \caption{
    Same as Figure~\ref{spectrum:b4.10} but at $\beta$ = 4.07
    ($T\sim$ 203~MeV).
  }
  \label{spectrum:b4.07}
\end{figure*}
\begin{figure*}[tbp]
    \centering
    \includegraphics[width=8.0cm]{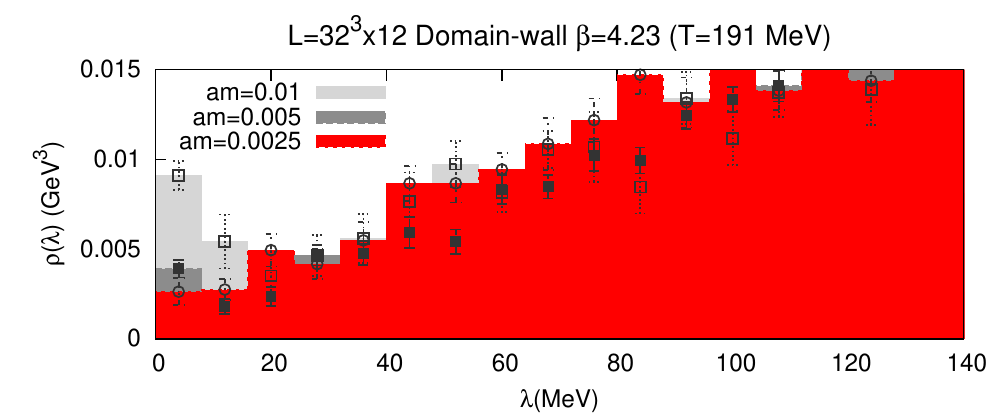}
    \includegraphics[width=8.0cm]{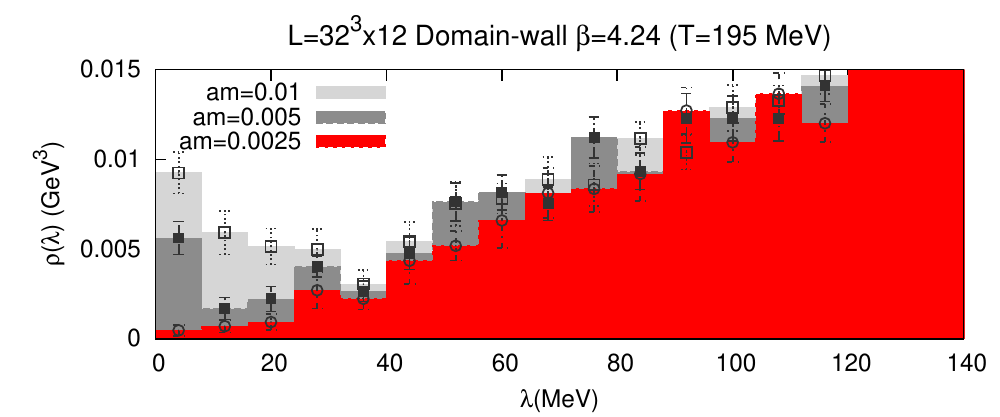}
    \includegraphics[width=8.0cm]{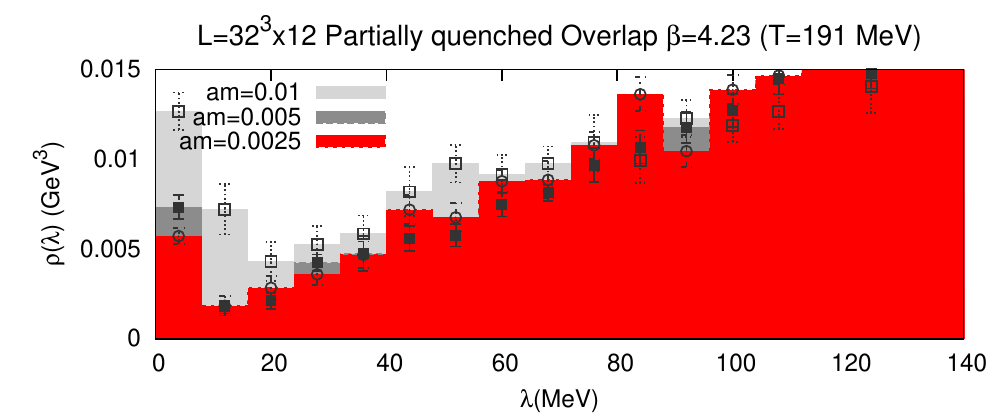}
    \includegraphics[width=8.0cm]{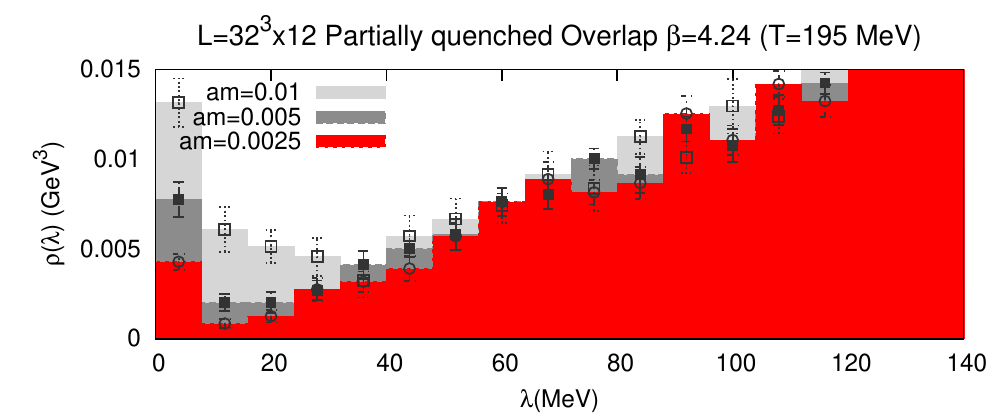}
    \includegraphics[width=8.0cm]{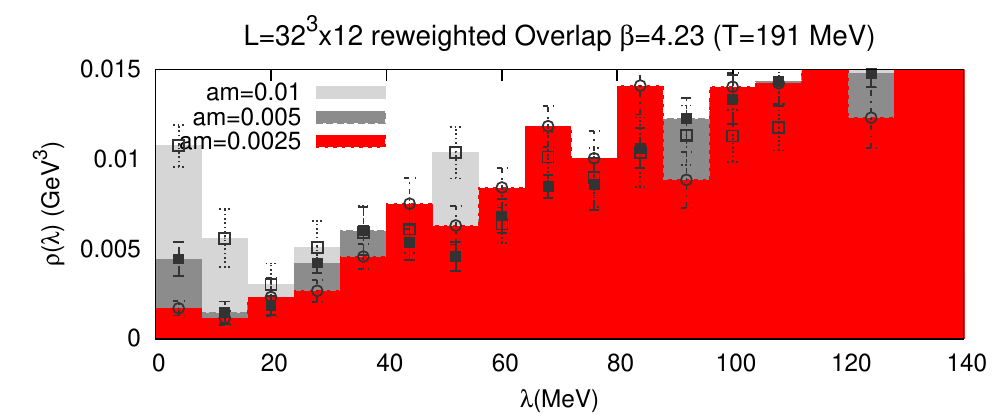}
    \includegraphics[width=8.0cm]{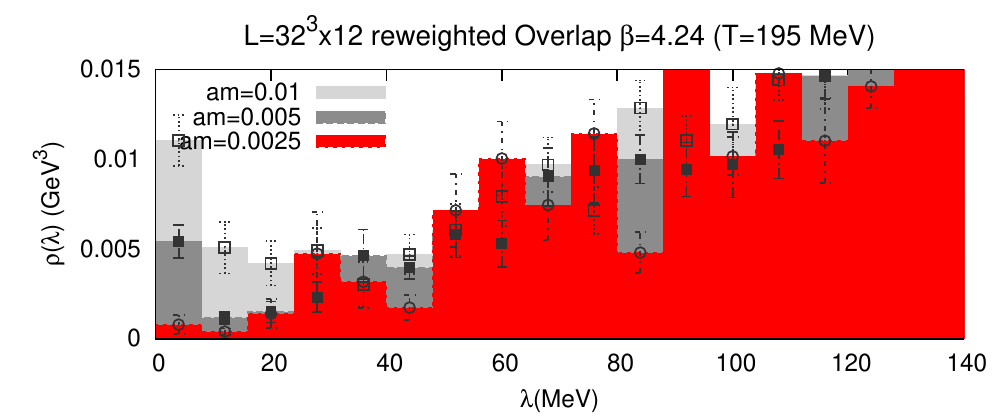}
  \caption{
    Same as Figure~\ref{spectrum:b4.10} but on finer lattices.
    Data at $\beta$ = 4.23
    ($T\sim$ 191~MeV)  (left panel) and 
    at  $\beta$ = 4.24 ($T\sim$ 195~MeV) (right) on the $32^3\times 12$ lattice are plotted.
  }
  \label{spectrum:higherbeta}
\end{figure*}

The above properties of the Dirac eigenvalue spectrum
are insensitive to the volume and lattice spacing,
as presented in Figs. \ref{spectrum:b4.07} and \ref{spectrum:higherbeta}.
We emphasize that the strong suppression 
of the near-zero modes does not change even when we vary
the lattice volume size from 2 fm to 4 fm, where the latter volume is 8 times larger than the former.
If there were a (pseudo) gap due to the finite volume,
it should scale as a power of $1/L$.
We also confirm that the screening mass 
in the pseudoscalar channel $M_{PS}$ is large enough to satisfy $M_{PS} L> 5$,
and the finite volume effects are well under control.

\begin{figure*}[tbp]
    \centering
    \includegraphics[width=8.0cm]{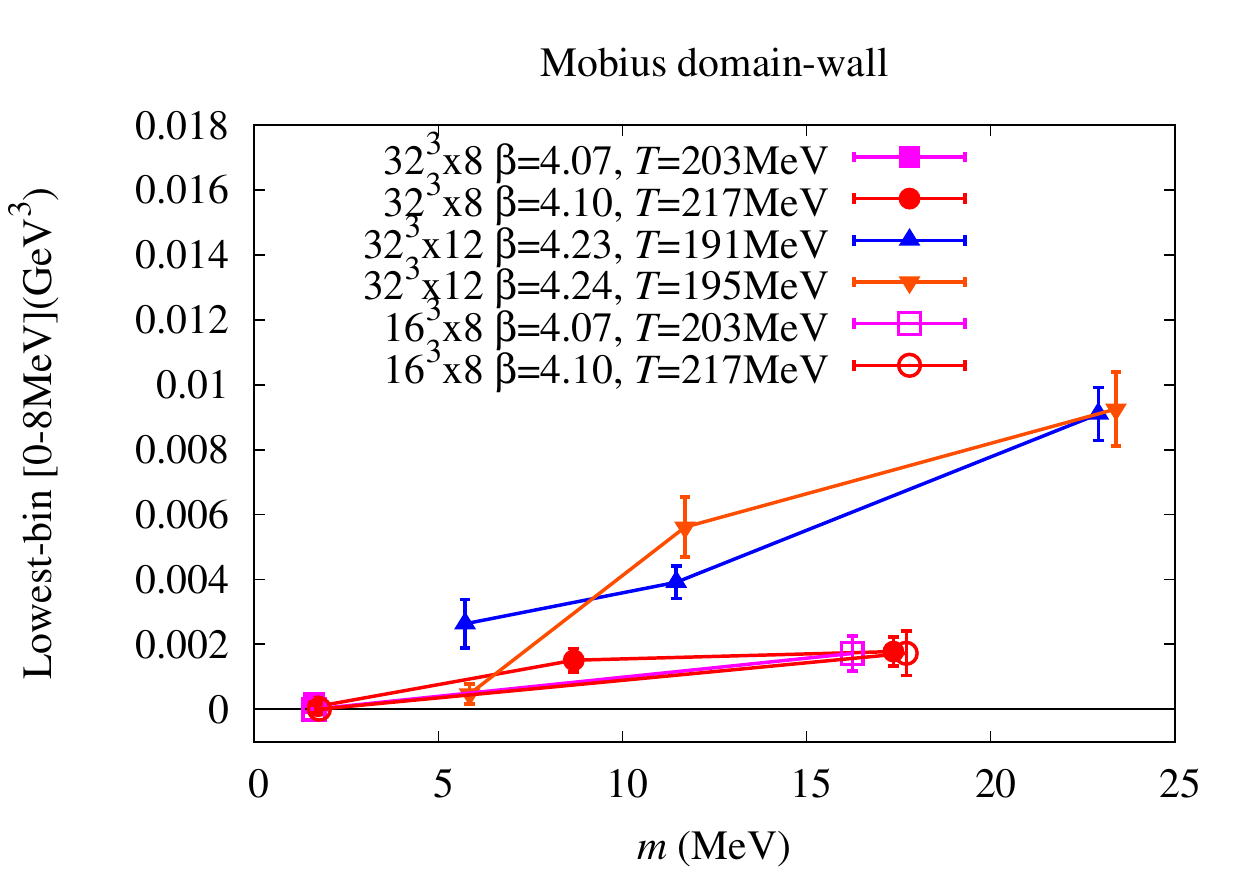}
    \includegraphics[width=8.0cm]{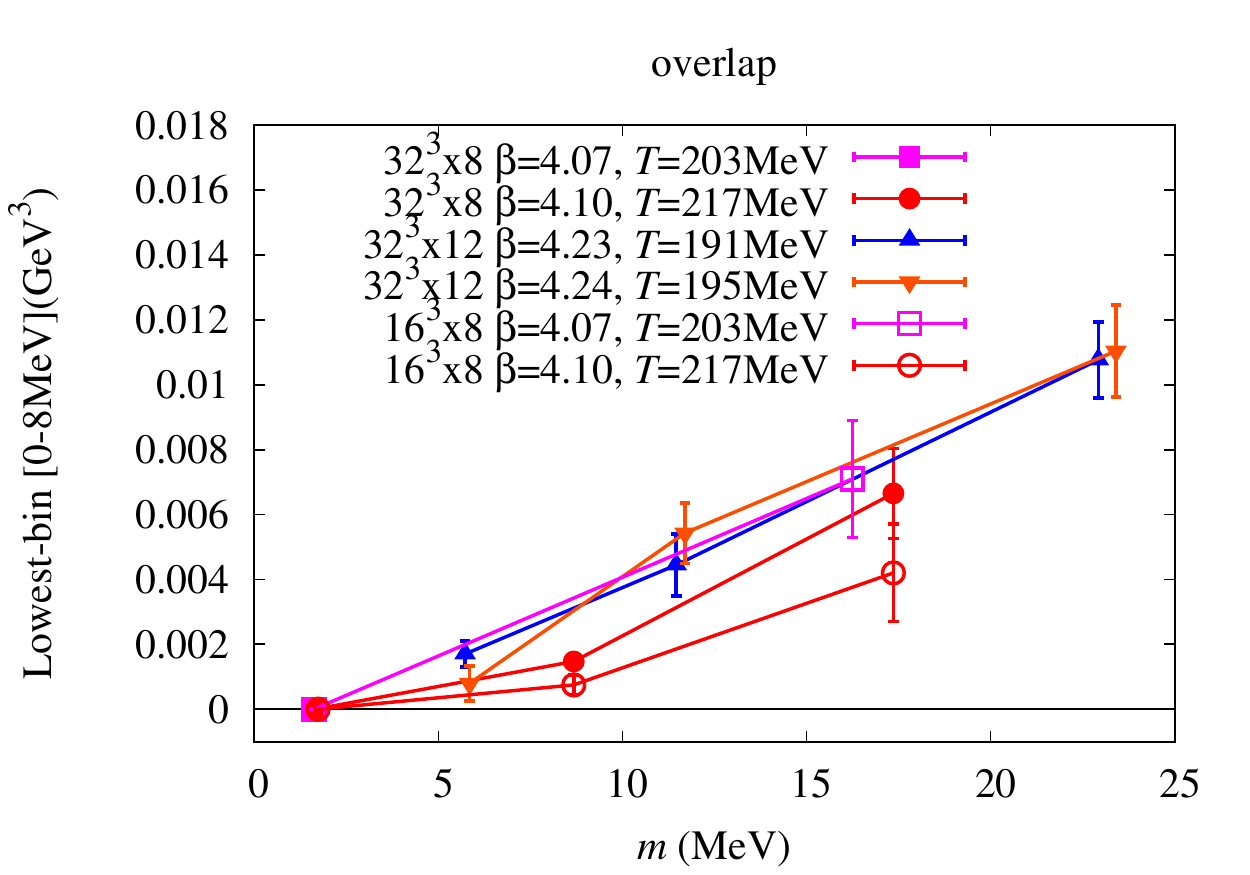}
  \caption{
    The quark mass dependence of the eigenvalue density 
    at the first bin [0,8] MeV. The data of the M\"obius domain-wall (left panel)
    and those of the overlap (right) Dirac operators are shown.
    All the data for $m<5$ MeV are consistent with zero.
  }
  \label{spectrum:m-dep1stbin}
\end{figure*}

We also remark that our data at the lowest bin, 
or the eigenvalue density below 8 MeV, 
show a monotonically decreasing quark mass dependence,
as presented in Fig.~\ref{spectrum:m-dep1stbin}.
Both data of the M\"obius and overlap Dirac operators 
at $m<5$ MeV are consistent with zero.

Since the difference between the \mobius domain-wall and
the reweighted overlap Dirac spectra are not clear,
one may think that they are qualitatively the same
and the overlap/domain-wall reweighting is not needed for the analysis
of the \ua susceptibility.
However, as we already reported in \cite{Cossu:2015kfa}\footnote{
The detailed study of the individual eigenmodes and their localization properties is 
also reported in Ref.~\cite{Cossu:2016scb}.
},
we found a significant difference in the chiral symmetry 
of the individual eigenmodes of the two Dirac operators.
In \cite{Cossu:2015kfa}, we investigated the effects of
chiral symmetry violation, or violation of the Ginsparg-Wilson relation on each eigenmode,
and how they affect the physical observables. 
For example, we observed
\begin{align}
g_i &\equiv
\frac{
\psi_i^\dagger\Delta_\text{GW}\psi_i}
{\lambda_i^{(m)}}
\left[
\frac{(1-am)^2}{(1+am)}
\right]
\label{eq:def_gi},
\end{align}
where $\Delta_\text{GW}$ is defined by (\ref{eq:def_DeltaGW}), and
$\lambda_i^{(m)}$, $\psi_i$ denote the $i$--th eigenvalue/eigenvector of massive hermitian Dirac operator, respectively.
The last factor in (\ref{eq:def_gi}) comes from the normalization of the Dirac operator.
$g_i$ vanishes when the \gw is exactly satisfied.

\begin{figure}[tbp]
  \centering
  \includegraphics[width=10cm]{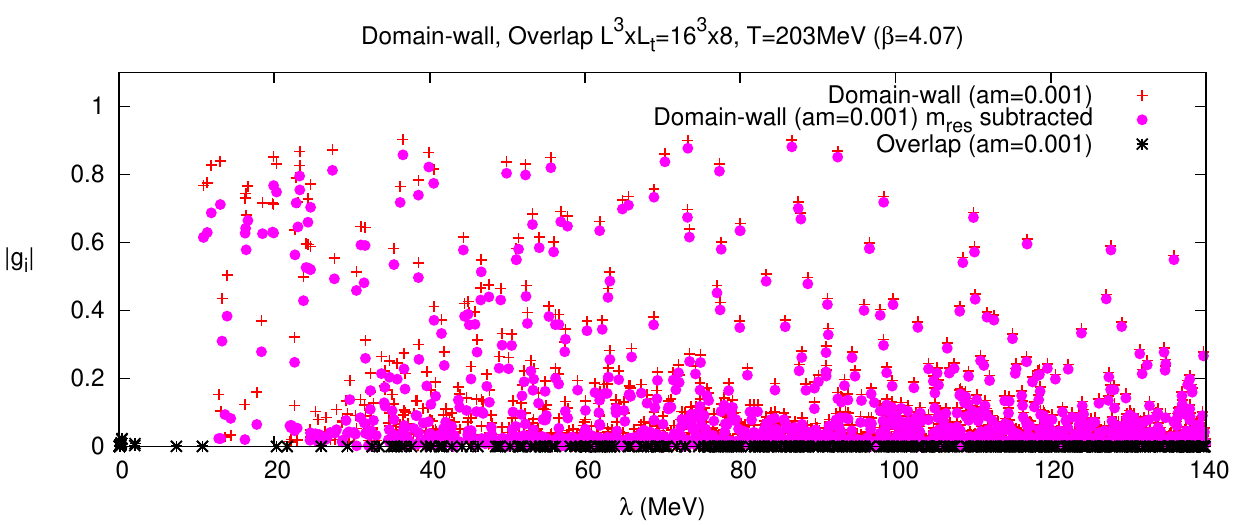}
  \includegraphics[width=10cm]{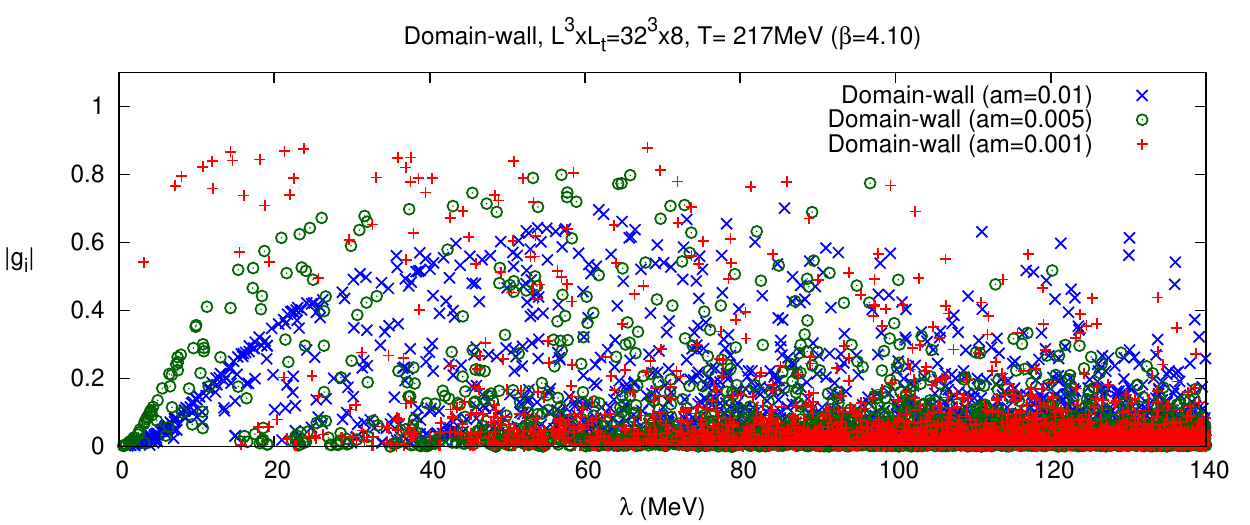}
  \includegraphics[width=10cm]{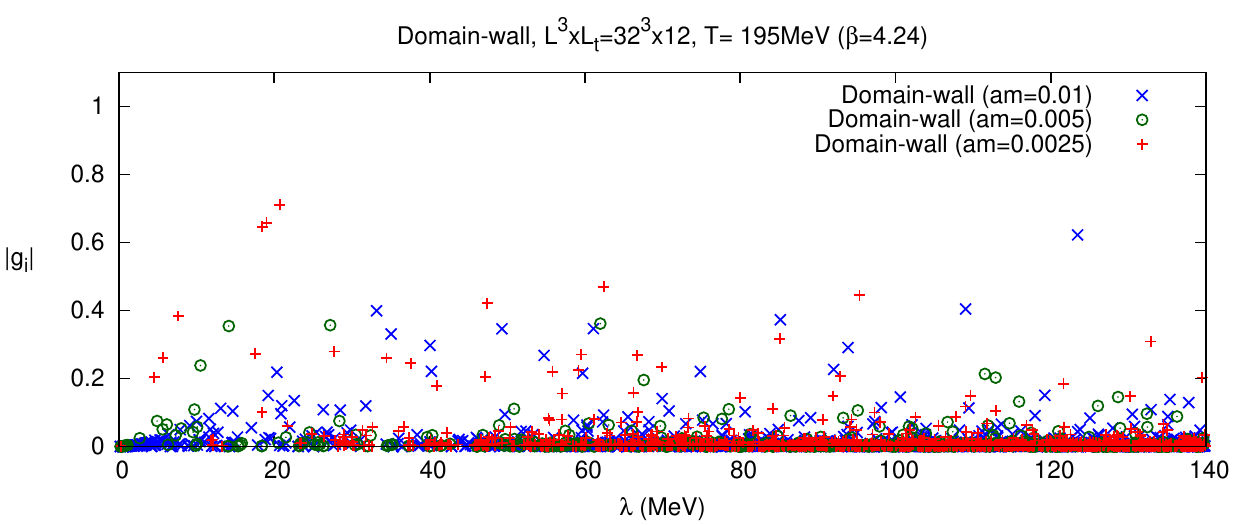}
\caption{Violation of the Ginsparg-Wilson relation $g_i$ as measured for individual eigenmodes.
 Data for $\beta=4.07$ on a $L^3\times L_t=16^3\times 8$ lattice (top panel),
 those for $\beta=4.10$ and  $L^3\times L_t=32^3\times 8$ (middle), and
 those for $\beta=4.24$ and  $L^3\times L_t=32^3\times 12$ (bottom) are shown.
 Results for all the measured configurations are plotted.
}
\label{fig:L16b407_dw_gi_ov.pdf}
\end{figure}

As shown in Fig.~\ref{fig:L16b407_dw_gi_ov.pdf} we
found that the low-lying modes of the \mobius domain-wall Dirac operator (cross symbols)
violate the chiral symmetry to the order of one, 
which means that the expectation value of the violation of the \gw
is comparable 
to the eigenvalue  $\lambda_i^{(m)}$ itself.
On the top panel of Fig.~\ref{fig:L16b407_dw_gi_ov.pdf}, 
we also plotted the data using $D^{4D}_\text{DW}(-m_\text{res})$ (solid circles) 
instead of $D^{4D}_\text{DW}(0)$,
expecting some cancellation with the effect from the residual mass\footnote{
We thank Y. Shamir for the suggestion for the improvement by subtracting $m_\text{res}$.
}.
The improvement is at most 20 \%, and the violation remains to be $\mathcal{O}(1)$.
For the overlap Dirac operator (star symbols on the top panel), 
$g_i$ is negligibly small, as expected.

This large violation of chiral symmetry may potentially
distort the physical observables, if they are sensitive to 
these low-lying modes and their chirality.
As will be shown in the next section (and it is discussed in detail in \cite{Cossu:2015kfa}),
we find that at our lightest simulated mass,  60-90\%  
of the \ua susceptibility measured by the \mobius domain-wall 
Dirac fermions comes from the violation of the \gw.

The violation of the \gw on the low-lying modes
of the \mobius domain-wall Dirac operator
explains the large difference between the partially quenched 
and the reweighted overlap fermions.
Not only the valence fermions but also sea fermions 
are required to satisfy a good chirality, otherwise
the physical observables can be largely distorted.
Therefore, the overlap/domain-wall reweighting is essential in our analysis.

\begin{table}[tbp]
\begin{center}
\begin{tabular}{c|c|c|c|c|c|c|c|c}
$L^3\times L_t$ & $\beta$ & $m$ &
$\rho_{\rm ov}(\mbox{0--8MeV})$ & $\Delta_{\pi-\delta}^{\rm direct}a^2$
& $\Delta_{\pi-\delta}^{\rm ev}a^2$ &
$\Delta_{\pi-\delta}^{\cancel{\rm GW}}/\Delta_{\pi-\delta}^{\rm ev}$ &
$\Delta_{\pi-\delta}^{\rm ov}a^2$
& $\bar{\Delta}_{\pi-\delta}^{\rm ov}a^2$
\\ \hline\hline
      $16^3\times 8$ &$4.07$& 0.01 & 0.0071(18)&0.132(14)&
      0.139(12)&0.37(2)&0.19(5)&0.032(13)\\
      $16^3\times 8$ &$4.07$& 0.001& 3(3)$\times 10^{-12}$ &0.032(7) &
      0.0498(14)&0.982(2)&0.00015(5)& 1.5(6)$\times 10^{-4}$\\

      $16^3\times 8$ &$4.10$& 0.01 & 0.0042(15)&0.073(12)&
      0.064(11)&0.278(40)&0.074(19)& 0.012(6)\\
      $16^3\times 8$ &$4.10$& 0.005$^*$& 0.0008(3) & 0.009(2)&
      --&--&0.0003(1) &0.003(1) \\
      $16^3\times 8$ &$4.10$& 0.001& 1.5(1.5)$\times
      10^{-8}$&0.017(8)& 0.0232(13)&0.983(4)&6(3)$\times 10^{-5}$&
      6(3)$\times 10^{-5}$\\

      $32^3\times 8$ &$4.07$& 0.001& 0.00002(1)& 0.105(32)&
      0.105(35)&0.65(10)& 0.03(2)&-0.004(3)\\

      $32^3\times 8$ &$4.10$& 0.01 & 0.0067(14) & 0.076(5) &
      0.069(5)&0.30(2) & 0.120(24)& 0.065(29)\\
      $32^3\times 8$ &$4.10$& 0.005& 0.00147(20) & 0.111(16)&
      0.107(15)&0.17(2)& 0.111(34)& 0.025(9)\\
      $32^3\times 8$ &$4.10$& 0.001& $1.5(1.3)\times10^{-5}$& 0.036(11)&
      0.0125(50)&0.975(3)& 0.097(38)& -0.010(5)\\

      $32^3\times 12$ &$4.23$& 0.01& 0.011(1)  & 0.112(10)&
      0.109(4)&0.038(4)& 0.11(1)&0.064(11)\\
      $32^3\times 12$ &$4.23$& 0.005& 0.00444 (96)  & 0.107(11)&
      0.107(8)&0.083(9)& 0.115(16)&0.026(7)\\
      $32^3\times 12$ &$4.23$& 0.0025& 0.0017(4)& 0.186(47)&
      0.216(41)&0.162(22)&0.162(40)&0.0065(20)\\

      $32^3\times 12$ &$4.24$& 0.01  & 0.011(1) & 0.135(8) &
      0.101(3)&0.046(3)& 0.107(14)&0.065(10)\\
      $32^3\times 12$ &$4.24$& 0.005 & 0.0054(9)&
      0.112(17)&0.124(13)&0.057(10)&0.122(21)& 0.030(14)\\
      $32^3\times 12$ &$4.24$& 0.0025& 0.0008(5)& 0.052(15)&
      0.041(13)&0.32(8) & 0.078(52)&0.0030(6)\\
      \hline
      \end{tabular}
      \caption{
      Summary of results. The data with the subscript ``ov''
      denote those with reweighted overlap fermions,
      otherwise, those with \mobius domain-wall fermions.
      \label{tab:results}
      The results at $\beta=4.10$, $m=0.005$ on the $16^3\times 8$ lattice
      (for which the asterisk is put)
      are obtained by choosing $m=0.005$ for the overlap Dirac operator to
      reweight the \mobius domain-wall ensemble generated with $m=0.01$.
      }
      \end{center}
      \end{table}



\section{\ua Susceptibility}
\label{sec:sus}

In this section, we directly investigate 
the \ua anomaly at high temperature by computing the susceptibility $\Delta_{\pi-\delta}$ in (\ref{eq:Delta_def}),
which is obtained from the two-point correlators in the iso-triplet scalar and pseudoscalar channels.
These correlators are related by the $U(1)_A$ symmetry and their difference must vanish
when the symmetry is recovered. The use of the iso-triplet channels has a practical advantage of not including disconnected diagrams,
which are numerically demanding.
In the previous work by JLQCD using dynamical overlap fermions \cite{Cossu:2013uua} we measured the meson correlators at finite temperature
and found that at temperatures close to the phase transition the mesons correlators coincide in the limit of small bare quark masses. 
Here we re-examine the \ua anomaly having a better control of systematic errors from
finite volumes and finite lattice spacings.

\begin{figure*}[tbp]
    \centering
    \includegraphics[width=8.0cm]{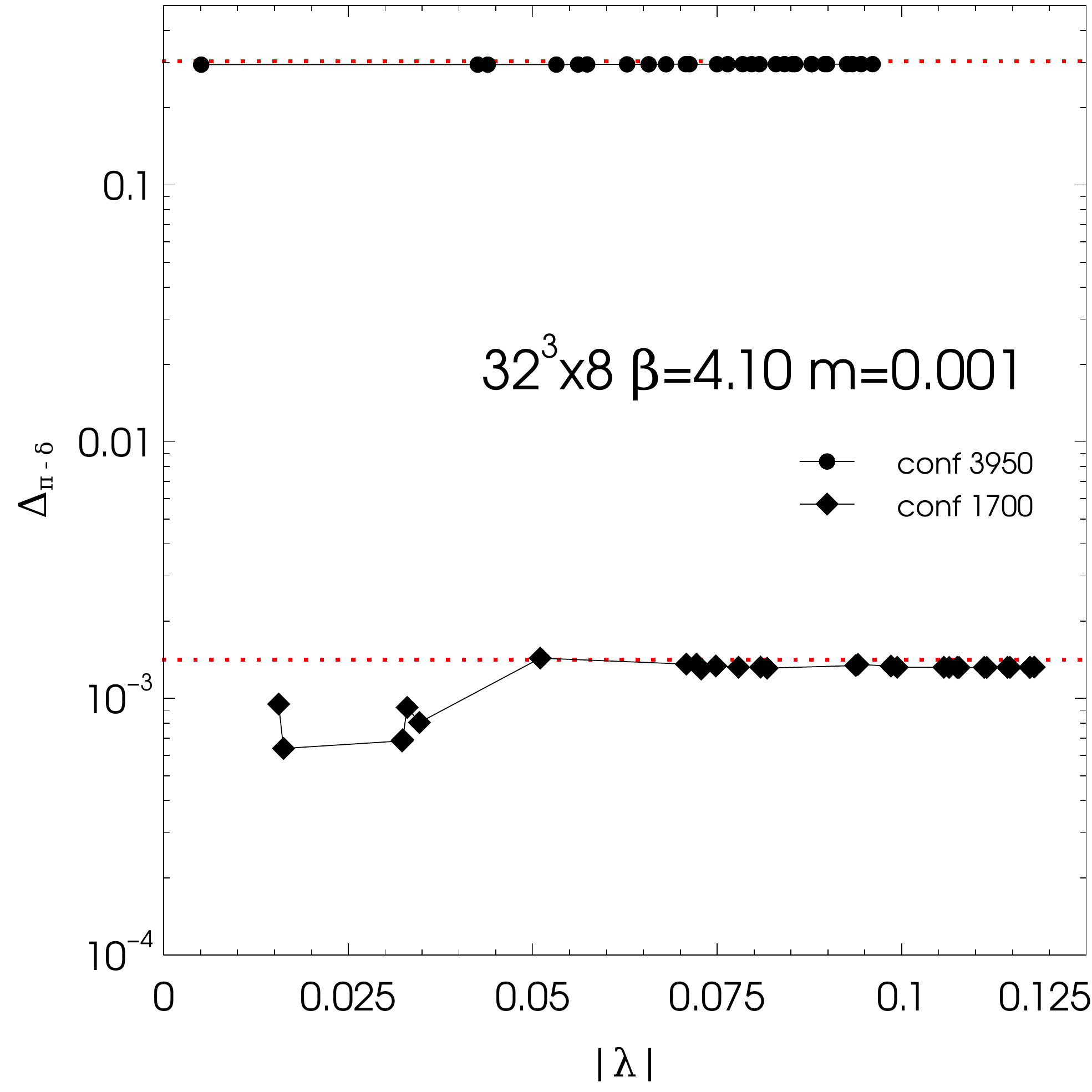}
  \caption{
  Low-mode saturation of $\Delta_{\pi-\delta}$. The horizontal axis shows the
  threshold of the eigenvalue, below which $\Delta_{\pi-\delta}^{\rm ev}$ is computed.
 The data for two typical configurations   
  generated with $\beta=4.10$, $ma=0.001$ on the $32^3\times 8$ lattice are shown. 
  The dotted lines are the results for the direct computation $\Delta_{\pi-\delta}^{\rm direct}$.
  }
  \label{fig:lowmodesaturation}
\end{figure*}

First, we examine how much the low-lying modes of 
the Dirac operator contribute to $\Delta_{\pi-\delta}$.
The strong violation of the chiral symmetry in the low-lying modes, 
found in the previous section, may affect the results.
We compare the eigenvalue decomposition of $\Delta_{\pi-\delta}$ (at finite $m$)
with $N_{\rm ev}=20$--100 lowest eigenmodes of the \mobius domain-wall Dirac operator,
$\Delta_{\pi-\delta}^{\rm ev}$,
and that directly computed by inverting the Dirac operator, $\Delta_{\pi-\delta}^{\rm direct}$, 
with a stochastic average of the source points.
This source point averaging is essential since our data at each single source point are noisy.
We find that the lowest modes below $\lambda\sim 0.1/a$
are enough to saturate the signals on all simulated ensembles.
The results for $\Delta_{\pi-\delta}^{\rm direct}$ and $\Delta_{\pi-\delta}^{\rm ev}$ are 
presented in Table~\ref{tab:results}.
The saturation of the low-mode approximation is demonstrated
in Fig.~\ref{fig:lowmodesaturation} for typical two configurations.

Next, let us separate the contribution coming from the violation of the \gw.
As we already discussed in \cite{Cossu:2015kfa},
$\Delta_{\pi-\delta}^{\rm ev}$ can be decomposed into the chiral symmetric part $\Delta_{\pi-\delta}^{\rm GW}$ 
and violating part $\Delta_{\pi-\delta}^{\cancel{\rm GW}}$ as
\begin{align}
\Delta_{\pi-\delta}^{\rm ev} &= \Delta_{\pi-\delta}^{\rm GW}+\Delta_{\pi-\delta}^{\cancel{\rm GW}},\\
\Delta_{\pi-\delta}^{\rm GW} &\equiv  \frac{1}{V(1-m^2)^2} \sum_i \frac{2m^2(1-\lambda_i^{(m)2})^2}{\lambda_i^{(m)4}},\\
\Delta_{\pi-\delta}^{\cancel{\rm GW}}&\equiv \frac{1}{V(1-m)^2} \sum_i \Bigl[\frac{h_{i}}{\lambda_i^{(m)}} - 4\frac{g_{i}}{\lambda_i^{(m)}}   \Bigr],
\end{align}
where $g_i$ was already defined in Eq.~(\ref{eq:def_gi}) and 
\begin{align}
h_i \equiv &
\frac{2(1-m)^2}{(1+m)}
\psi_i^\dagger\gamma_5 (H_\text{DW}^\text{4D}(m))^{-1}\gamma_5\Delta_\text{GW}
(H_\text{DW}^\text{4D}(m))^{-1}\psi_i 
+\frac{2}{1+m}\left(1+\frac{m}{\lambda_i^{(m)2}}\right)g_i
\label{eq:def_hi},
\end{align}
is another measure of the violation of \gw.
Both of these quantities must be zero if the \gw is satisfied. 

\begin{figure*}[tbp]
    \centering
    \includegraphics[width=8.0cm]{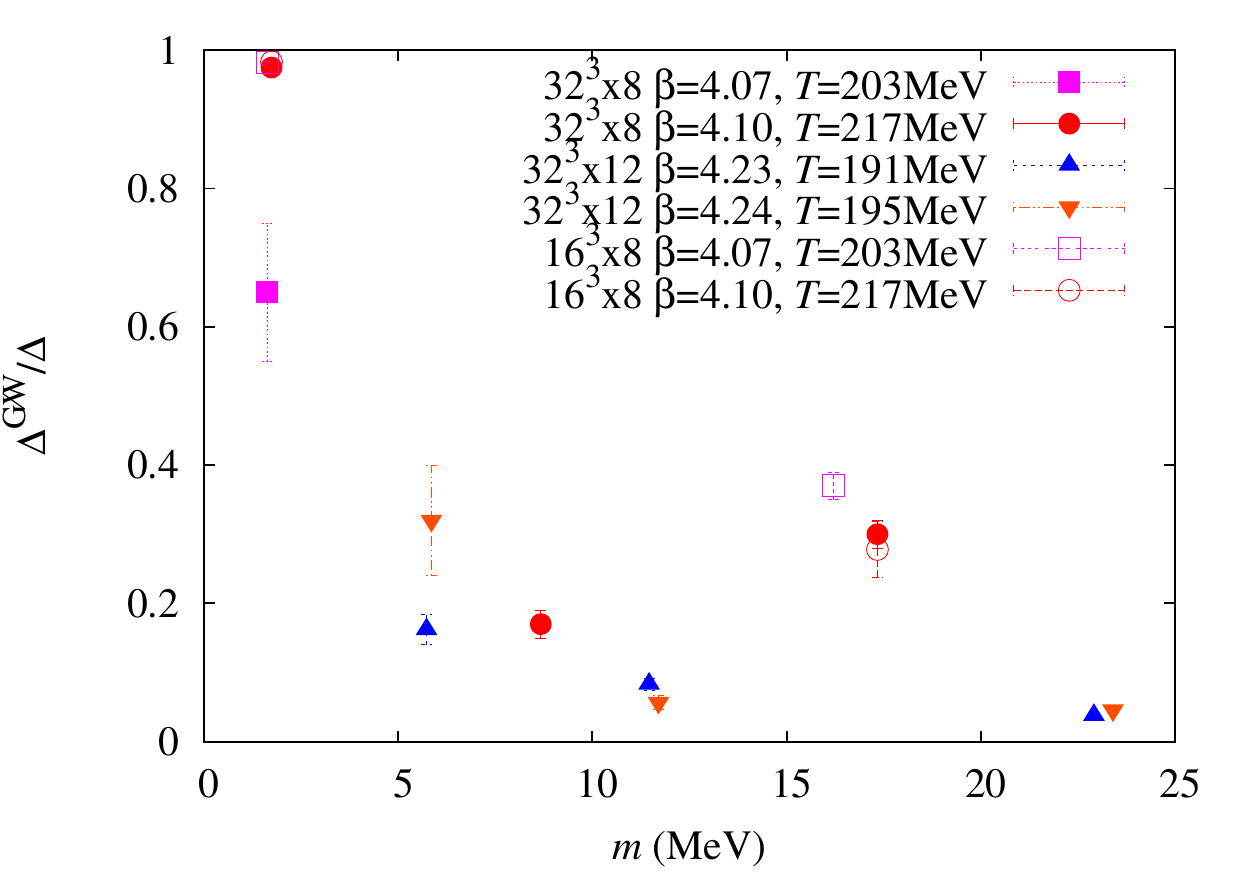}
  \caption{
  Quark mass dependence of the ratio $\Delta_{\pi-\delta}^{\cancel{\rm GW}}/\Delta_{\pi-\delta}$.
  The contribution from the chirality violating terms dominates the signal
  near the chiral limit.
  }
  \label{fig:Delta-DW-GWviolation}
\end{figure*}

Figure~\ref{fig:Delta-DW-GWviolation} shows the quark mass dependence of 
the ratio $\Delta_{\pi-\delta}^{\cancel{\rm GW}}/\Delta_{\pi-\delta}^{\rm ev}$.
The \gw violating part $\Delta_{\pi-\delta}^{\cancel{\rm GW}}$ 
dominates the signal as decreasing the quark mass.
For data points less than $m=5$ MeV (at lower $\beta$), more than 60--98 \%
of the signal is the contribution from $\Delta_{\pi-\delta}^{\cancel{\rm GW}}$.
Thus, we need a careful control of the chiral symmetry on the
low-lying eigenmodes in taking the chiral limit of the \ua breaking observables.

Finally, let us examine the \ua susceptibility with overlap fermions.
Here we do not use the partially quenched overlap as we have
shown its significant lattice artifacts.
We observe that the partially quenched overlap $\Delta_{\pi-\delta}$ overshoots the \mobius domain-wall data.
We confirm that $g_i$ and $h_i$ for the overlap Dirac eigenmodes 
are negligible (see Fig.~\ref{fig:L16b407_dw_gi_ov.pdf}),
so that we can safely use $\Delta_{\pi-\delta}^{\rm GW}$ together with the OV/DW reweighting to estimate 
the \ua susceptibility (let us denote it as $\Delta_{\pi-\delta}^{\rm ov}$).

\begin{figure*}[tbp]
    \centering
    \includegraphics[width=8.0cm]{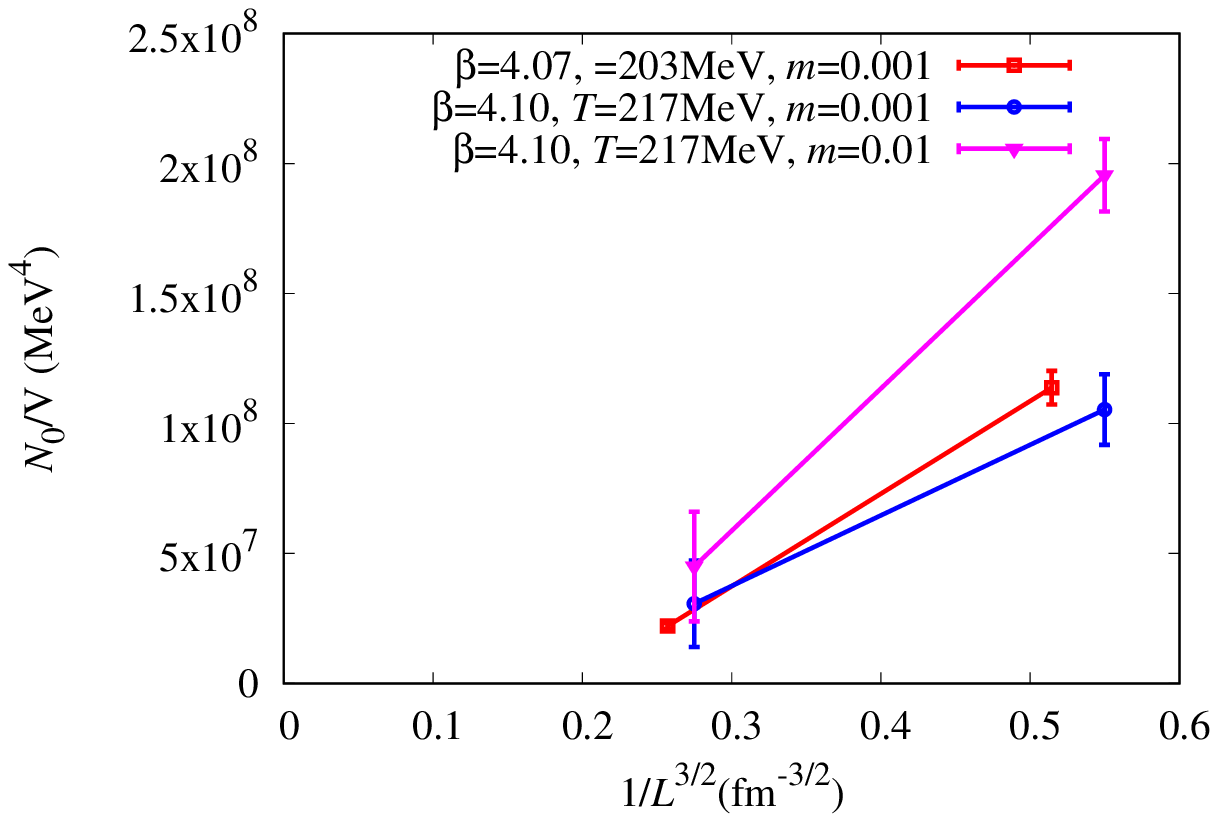}
  \caption{
  The lattice size $L$ dependence of $\langle N_0/V\rangle$.
  The results at $L_t=8$ are shown.
  }
  \label{fig:N0}
\end{figure*}

Taking the advantage of good chirality, 
we can subtract the effect of the chiral zero-mode effects\footnote{
   The eigenmode decomposition is given in Eq.~(3.11) of our previous work \cite{Cossu:2015kfa},
   from which we can identify the zero-mode contribution is $\frac{2N_0}{Vm^2}$.
}:
\begin{align}
\label{eq:Deltabar}
\bar{\Delta}_{\pi-\delta}^{\rm ov} \equiv \Delta_{\pi-\delta}^{\rm ov} - \frac{2N_0}{Vm^2}.
\end{align}
The expectation value of $N_0^2$ is expected to be an $O(V)$ quantity, 
as shown in \cite{Aoki:2012yj},
so that these chiral zero-mode's effects should not survive in the large volume limit,
as $N_0/V$ is vanishing as $O(1/\sqrt V)$.
We numerically confirm  the monotonically decreasing volume scaling of 
$\langle N_0/V\rangle$ as shown in Fig.~\ref{fig:N0}.
Therefore, $\bar{\Delta}_{\pi-\delta}^{\rm ov}$ and $\Delta_{\pi-\delta}^{\rm ov}$ are guaranteed 
to have the same thermodynamical limit.
We also confirm that the 5--15 lowest modes are enough to saturate the
reweighing for  $\bar{\Delta}_{\pi-\delta}^{\rm ov}$ on $32^3\times 8$ lattices.

\begin{figure*}[tbp]
    \centering
    \includegraphics[width=8.0cm]{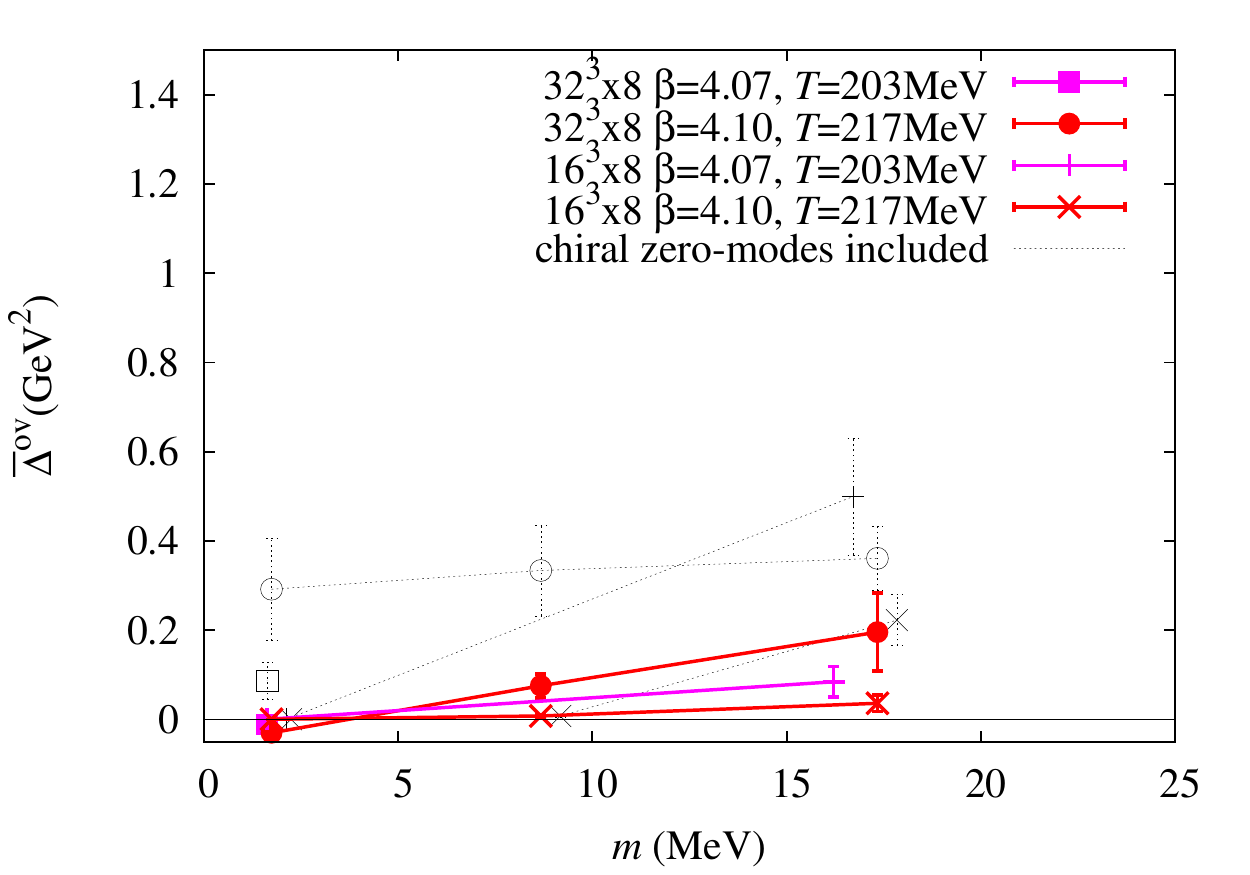}
    \includegraphics[width=8.0cm]{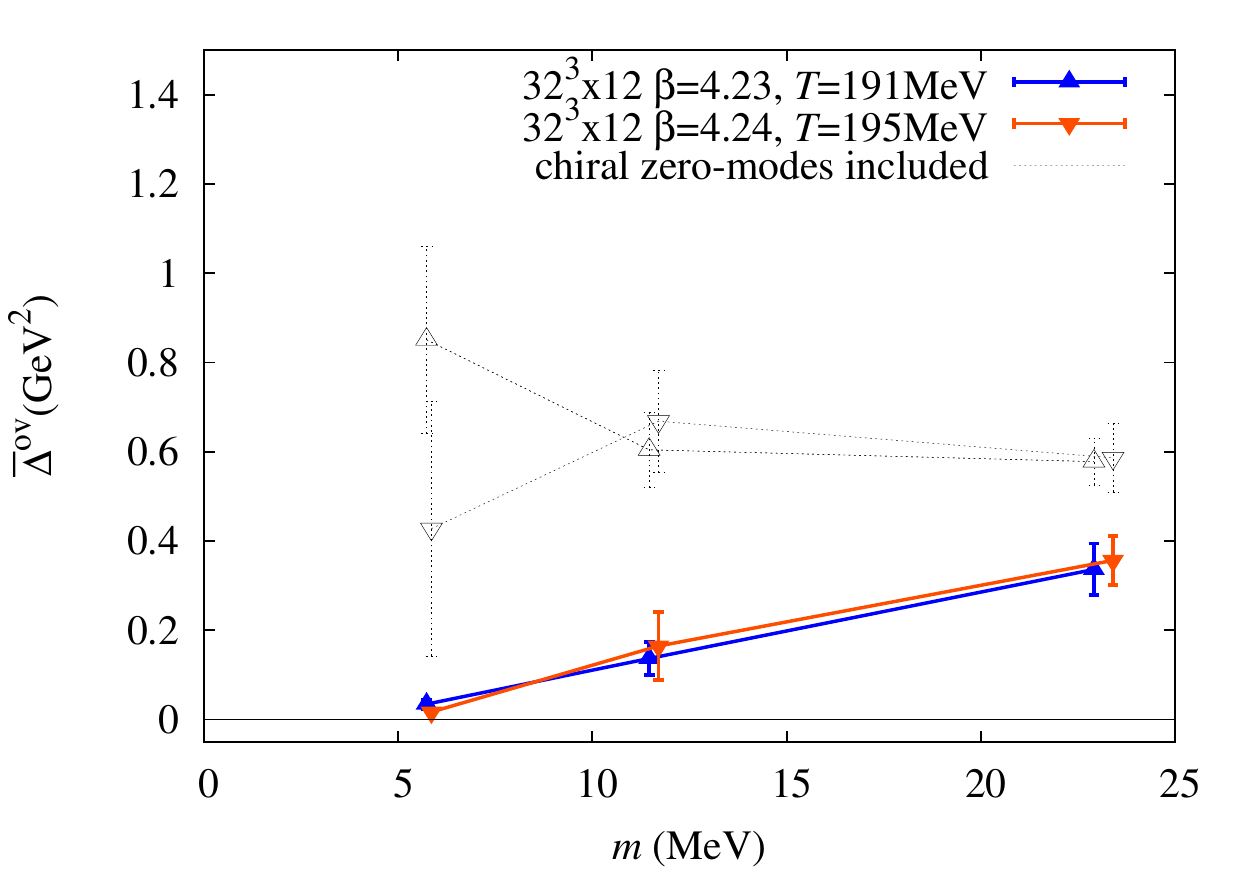}
  \caption{
  The quark mass dependence of $\bar{\Delta}_{\pi-\delta}^{\rm ov}$ (solid symbols) and $\Delta_{\pi-\delta}^{\rm ov}$ (dashed).
  Data for coarse (left panel) and fine (right) lattices are shown.
  }
  \label{fig:FinalAfterRW}
\end{figure*}

Our results for $\bar{\Delta}_{\pi-\delta}^{\rm ov}$ (solid symbols)
and $\Delta_{\pi-\delta}^{\rm ov}$ (dashed)
are plotted in Fig.~\ref{fig:FinalAfterRW}.
We confirm that our data for $\bar{\Delta}_{\pi-\delta}^{\rm ov}$ are
stable against
the change of the lattice size, and lattice spacing,
and their chiral limits are all consistent with zero.
Precisely, all our data are  well described (with $\chi^2$/d.o.f $\lesssim 1$)
by a simple linear function, which becomes consistent with zero
``before'' the chiral limit.
We list the linear extrapolation of $\bar{\Delta}_{\pi-\delta}^{\rm
ov}$ at $m_\text{ud}=4$ MeV\footnote{
Because we don't renormalize the quark mass,
our data do not represent physically equal mass points.
However, their difference is expected to be smaller than
the statistical errors and systematic errors of the fit.
The conclusion does not change if we choose $m_\text{ud}<5$ MeV.
} in Table~\ref{tab:4MeVlimit}.
We observe 
neither strong volume dependence nor  $\beta$ dependence of this behavior.
Taking the largest value in the table, we conclude that the chiral limit of
$\bar{\Delta}_{\pi-\delta}^{\rm ov}$ is estimated to be at most
0.0040(130) GeV$^2$.
Although our naive linear extrapolation may simply fail to detect
higher order mass dependence,
the smallness of $\bar{\Delta}_{\pi-\delta}^{\rm ov}$ itself compared
to the data around $m_\text{ud}=20$ MeV is significative
and has a phenomenological importance.

\begin{table}[tbp]
\begin{center}
\begin{tabular}{ccc|cc}
$L^3\times L_t$ & $\beta$ & $T$(MeV) &$\bar{\Delta}_{\pi-\delta}^{\rm
ov}$[GeV$^2$] at $m_\text{ud}=4$ MeV　& $\chi^2$/d.o.f.
\\ \hline\hline
      $32^3\times 12$ &$4.23$& 191(1) & 0.0037(099) & 0.002\\
      $32^3\times 12$ &$4.24$& 195(1) & -0.0199(033) & 0.2\\
      $16^3\times 8$ &$4.10$& 217(1) & 0.0025(017) & 1.0\\
      $32^3\times 8$ &$4.10$& 217(1) & 0.0040(130) & 0.01\\
      \hline
\end{tabular}
      \caption{
        Linear extrapolation of $\bar{\Delta}_{\pi-\delta}^{\rm ov}$
to $m_\text{md}=4$ MeV.
        It becomes consistent with zero before the chiral limit.
        \label{tab:4MeVlimit}
      }
\end{center}
\end{table}

\section{Conclusion}
\label{sec:conclusion}
In this work, we have examined the \ua anomaly in two-flavor 
lattice QCD at finite temperature with chiral fermions.
On the configurations generated by the \mobius domain-wall Dirac quarks,
we have measured the Dirac eigenvalue spectrum both of the \mobius domain-wall and overlap quarks,
with or without OV/DW reweighting. 
We have also examined the meson susceptibility difference $\Delta_{\pi-\delta}$, 
that directly measures the violation of the \ua symmetry.
Our ensembles are generated at slightly above the critical temperature of 
the chiral phase transition ($T\sim 190$--220~MeV)
on different physical volume sizes ($L=$ 2--4 fm), where frequent topology tunnelings occur.

Our results for the histograms of the \mobius domain-wall and  (reweighted) overlap 
Dirac operators both show a strong suppression of the near zero modes 
as decreasing the quark mass. 
This behavior is stable against the change of the lattice size and lattice spacing.

If we do not perform the reweighting of their determinants,
the overlap Dirac spectrum shows unphysical peaks near zero.
We have identified them as partially quenched lattice artifacts,
due to the strong violation of the \gw in the low-lying eigenmodes
of the \mobius domain-wall operator.
Our analysis indicates a potential danger in taking the chiral limit
of any observables with domain-wall type fermions even when the residual mass is small.
If the target observable is sensitive to the low-lying modes and their chiral properties,
its chiral limit can be distorted by the lattice artifacts.

After removal of these artifacts by the OV/DW reweighting procedure,
we have found that the \ua susceptibility is consistent with zero in the chiral limit.
From these evidences, 
we conclude that \ua symmetry breaking
in two-flavor QCD is consistent with zero above the critical
temperature around 200 MeV, in the vanishing quark mass limit.

\begin{acknowledgments}
We thank K.~Hashimoto, K.~Kanaya, T.~Kanazawa, Y.~Taniguchi for discussions.
We also thank the members of JLQCD collaboration for their support on this work.
Discussions during the YITP workshop YITP-T-14-03 on Hadrons and Hadron Interactions in QCD” 
were helpful to complete this work.
AT received generous support from H-T.~Ding.
Numerical simulations are performed on IBM System Blue Gene Solution at High Energy Accelerator Research Organization (KEK) under a support for is Large Scale Simulation Program (No. 14/15-10). 
This research was supported by MEXT as “Priority Issue on Post-K computer” (Elucidation of the Fundamental Laws and Evolution of the Universe).
This work is supported in part by
JSPS KAKENHI (Grant Number JP25800147, JP26247043, JP26400259, and JP15K05065, JP16H03978), and by MEXT SPIRE and JICFuS.
GC is supported by STFC, grant ST/L000458/1.
AT is supported by NSFC under grant no. 11535012.
\end{acknowledgments}


\end{document}